\documentclass[12pt,a4paper]{article}
\usepackage[usenames]{color}
\usepackage{graphicx}
\usepackage{rotating}
\usepackage{geometry}
\usepackage{pdflscape}
\usepackage{float}
\usepackage{fancyhdr}
\usepackage{amsmath,amssymb}
\usepackage{url}

\def\bm#1{\hbox{\boldmath  $#1$\unboldmath}{}}  
\def\bm#1{\hbox{\boldmath  $#1$\unboldmath}{}}
\newcommand{\ignore}[1]{}{}

\def\d{\color{black}}

\begin{document}


\title{Risk Margin Quantile Function Via Parametric and Non-Parametric Bayesian Quantile Regression}
\author{Alice X.D.~Dong$^{1}$ \quad Jennifer S.K.~Chan$^{1}$ \quad Gareth W.~Peters$^{2}$}
\date{{\footnotesize {Working paper, version from \today }}}
\maketitle
\par \vspace{-12mm}
\begin{center}
{\footnotesize {\ \textit{
$^{1}$ School of Mathematics and Statistics\\
The University of Sydney, NSW 2006, Australia\\[0pt]
email: xdon0433@uni.sydney.edu.au, (Corresponding Author) \\[0pt]
$^{2}$ Department of Statistical Science,University College London UCL, London, UK;\\[0pt]
} } }
\end{center}

\begin{abstract}

\noindent We develop quantile regression models in order to derive risk margin and to evaluate capital in non-life insurance applications. By utilizing the entire range of conditional quantile functions, especially higher quantile levels, we detail how quantile regression is capable of providing an accurate estimation of risk margin and an overview of implied capital based on the historical volatility of a general insurers loss portfolio. Two modelling frameworks are considered based around parametric and nonparametric quantile regression models which we develop specifically in this insurance setting.

In the parametric quantile regression framework, several models including the flexible generalized beta distribution family, asymmetric Laplace (AL) distribution and power Pareto distribution are considered under a Bayesian regression framework. The Bayesian posterior quantile regression models in each case are studied via Markov chain Monte Carlo (MCMC) sampling strategies.

In the nonparametric quantile regression framework, that we contrast to the parametric Bayesian models, we adopted an AL distribution as a proxy and together with the parametric AL model, we expressed the solution as a scale mixture of uniform distributions to facilitate implementation. The models are extended to adopt dynamic mean, variance and skewness and applied to analyze two real loss reserve data sets to perform inference and discuss interesting features of quantile regression for risk margin calculations.
\end{abstract}

Asymmetric Laplace distribution, Bayesian inference, Markov chain Monte Carlo methods, Quantile regression, loss reserve, risk margin, central estimate.



\section{Background on Risk Margin Calculation} \par \vspace{-4mm} \noindent
A core component of the work performed by general insurance actuaries involves the assessment, analysis and evaluation of the  uncertainty involved in the claim process with a view to assessing appropriate risk  margins for inclusion in insurance liabilities. An appropriate valuation of insurance liabilities including risk margin is one of the most important issues for a general insurer. Risk margin is the component of the value of claims liability that relates to the inherent uncertainty.

The significance of this task is well understood by the actuarial profession and has been debated by both practitioners and academic actuaries alike. Much of the attention involves the non prescriptive nature of risk margin requirements discussed in regulatory guidelines such as Article 77 and Article 101 of the Solvency II Directives. In Australia a general task force was established, developing a report on risk margin evaluation methodologies presented to the Australian actuarial profession at the Institute of  Actuaries of Australia during the 16-th General Insurance Seminar in 2008. This report aimed to highlight approaches to risk margin calculations that are often considered. Before briefly discussing these aspects we first note the following Solvency II items which relate to the Solvency Capital Requirement (SCR) and the risk margin.

\noindent Article 101 of the Solvency II Directive states,
\begin{center}
``\itshape{
The Solvency Capital Requirement (SCR) shall correspond to the Value-at-Risk (VaR) of the basic own funds of an insurance or reinsurance undertaking subject to a confidence level of 99.5\% over a one-year period.
}
''
\end{center}
Essentially, the basic own funds are defined as the excess of assets over liabilities, under specific valuation rules. In this regard, a core challenge is the capital market-consistent value of insurance liabilities, which requires a best estimate typically defined as the expected present value of future cash flows under Solvency II plus a risk margin calculated using a cost-of-capital approach.

\noindent Furthermore, under Article 77 of the 2009 Solvency II Directive it states that the risk margin calculation is described as
\begin{center}
``
\itshape{The risk margin shall be such as to ensure that the value of the technical provisions is equivalent to the amount insurance undertakings would be expected to require in order to take over and meet the insurance obligations...
… it shall be calculated by determining the cost of providing an amount of eligible own funds equal to the Solvency Capital Requirement necessary to support the insurance obligations over the lifetime thereof...
}
''
\end{center}

As can be seen from such specifications, the recommendations to be adopted are not prescriptive in the required model approaches. Therefore, as discussed in the white paper produced by the Risk Margins Taskforce 1998, there have been several approaches considered which range from those that involve little analysis of the underlying claim portfolio to those that  involve significant analysis of the uncertainty using a wide range of information and techniques, including stochastic modelling. They highlighted approaches adopted in practice in the assessment of risk margins and pointed to percentile or quantile methods as being most prevalent in practice, this provides a good foundation for the methods we consider.

Traditionally, actuaries that adopt a stochastic framework would evaluate claims liability using a central estimate which is typically defined as the expected value over the entire range of outcomes. However with the inherent uncertainty that may arise from such an estimator which is not statistically robust and therefore sensitive to outlier claims, claims liability measures often differ from their central estimates. In practice, the approach adopted is typically to then set an insurance provision so that, to a specified probability, the provision will eventually be sufficient to cover the run-off claims. For instance, in order to satisfy the requirement of the Australian Prudential Regulation Authority (APRA) to provide sufficient provision at a 75\%  probability level, the risk margin should be modelled statistically so that it can capture the inherent uncertainty of the mean estimate. When this margin is then added to the central estimate, it should provide a reasonable valuation of claims liability and therefore increases the likelihood of providing sufficient provision to meet the level required in GPS 320.  In this regard, it is worth noting that the more volatile a portfolios runoffs or those that display heavy tailed features may require a higher risk margin, since the potential for large swings in reserves is greater than that of a more stable portfolio.

To accommodate these ideas, two common methods for risk margin estimation have been proposed in practice. These are the cost of capital and the percentile methods. Under the cost of capital method the actuary determines the risk margin by measuring the return on the capital required to protect against adverse development of those unpaid claim liabilities. It is evident that application of the cost of capital method requires an estimate of the initial capital to support the unpaid claim liabilities and also the estimate of return on that capital. Alternatively, under the percentile or quantile method that we consider in this paper, which is currently used in Australia the actuary takes the perspective that the insurer must be able to meet its liability with some probability under some assumptions on the distribution of liabilities.  Risk margin is then calculated by subtracting the central estimate from a predefined critical percentile value.

What we bring to the percentile and quantile based framework in our proposed methods is the ability to incorporate in a rigorous statistical manner, regression factors that may be related to both exogenous features directly related to the insurance claims run-off stochastic process as well as endogenous factors that are related to for instance the current micro or macro economic conditions and the regulatory environment. These will be incorporated into a statistical model that allows one to explain the proportion of variation in the risk margin that is attributed to such features in a principled manner, as we shall demonstrate allowing for accurate estimation and prediction. We argue that since the percentile-based method involves the estimation of quantiles, it is therefore somewhat natural to consider quantile regression, which is a statistical technique to estimate conditional quantile functions, which can be used to estimate risk margin.

Just as classical linear regression methods based on minimizing sums of squared residuals enable one to estimate models for conditional mean functions, quantile regression methods offer a mechanism for estimating models for the conditional median function, and the full range of other conditional quantile functions. This model allows studying the effect of explanatory variables on the entire conditional distribution of the response variable and not only on its center. Hence we may develop factors and covariates which are explanatory of the risk margin variation directly through the proposed quantile regression framework. By supplementing the estimation of conditional mean functions with techniques for estimating an entire family of conditional quantile functions, quantile regression is capable of providing a more complete statistical analysis of the stochastic relationships among random variables.

Quantile regression has been applied to a wide range of applications in economics and finance, but has not yet been developed in a claim reserving context for risk margin estimation. We will demonstrate the features of quantile regression that have been popularized in finance and explain how they can be adopted in important applications in insurance, such as risk margin calculations. In quantitative investment, least square regression-based analysis is extensively used in analyzing factor performance, assessing the relative attractiveness of different firms, and monitoring the risks in their portfolios. Engle and Manganelli (2004) consider the quantile regression for the Value at Risk (VaR) model. They construct a conditional autoregressive value at risk model (CAVaR), and employ quantile regression for the estimation. The risk measure, VaR is defined as a quantile of the loss distribution of a portfolio within a given time period and a confidence level. Accurate VaR estimation can help financial institutions maintain appropriate capital levels to cover the risk from the corresponding portfolio.

Taylor (2006) estimate percentile-based risk margins via a parametric model based on the assumption of a log normal distribution of liability.  Other sophisticated distributions to capture flexible shapes and tail behaviors are also proposed to model severity distribution on aggregated claim data. These distributions include the generalized-$t$ (McDonald and Newey, 1988), Pareto (Embrechts {\em et al.}, 1997), the Stable family (Paulson and Faris, 1985; Peters, Byrnes and Shevchenko 2011; Peters, Shevchenko, Young and Yip, 2011), the Pearson family (Aiuppa, 1988), the log-gamma and lognormal (Ramlau and Hansen, 1988) and the lognormal and Burr 12 (Cummins {\em et al.}, 1999), and type II generalized beta (GB2) distribution (Cummins {\em et al.}, 1990, 1999, 2007). While these distributions on real support are flexible to model both leptokurtic and platykurtic data, they require log-transformation for claims data and the resulting log-linear model may be more sensitive to low values than large values (Chan {\em et al.}, 2008).

In Peters, Wuethrich and Shevchenko (2009) they adopt a Poisson-Tweedie family of models which incorporates families such as normal, compound poisson Gamma, positive stable and extreme stable distributions into a family of models. It was shown how such a generalized regression structure could be used in a claims reserving setting to model the claims process whilst incorporating covariate structures from the loss reserving structure. In this instance a multiplicative structure for the mean and variance functions was considered and quantiles were derived from modelling the entire distribution, rather than specifically targeting a model at the conditional quantile function.

Recently, in Dong and Chan (2013) an alternative class of flexible skew and heavy tail models was considered involving the GB2 distribution with positive support adopting dynamic mean functions and mixture model representation to model long tail loss reserving data and showed that GB2 outperforms some conventional distributions such as Gamma and generalised Gamma. The GB2 distribution family is very flexible as it includes both heavy-tailed and light-tailed severity distributions, such as gamma, Weibull, Pareto, Burr12, lognormal and the Pearson family, hence providing convenient functional forms to model claims liability. From the perspective of quantile specific regression models, recently Cai (2010) proposed a power-Pareto model which allows for flexible quantile functions which can provide a combination of quantile functions for both power and Pareto distributions. These combinations enable the modelling of both the main body and tails of a distribution.

The difference with our current methodology is that instead of developing a statistical model to capture all features of the claims run-off stochastic structure, with the incorporation of regression components, we propose, in this work, to target explicitly the conditional quantile functions in a regression structure. From a statistical perspective, this is a fundamentally different approach to these previously mentioned reserving model approaches. However we will illustrate that we can borrow from such models in developing our risk margin quantile regression framework. In fact the associate parameter estimation loss functions, parameter estimator properties and the resulting quantile in sample and out of sample forecasts will significantly differ to those achieved when trying to develop a model for the entire process rather than targeting the quantity of interest in this case, the particular quantile level. This is clear from the perspective that only under a Gaussian distributional assumption for such reserve models (on log scale) would a standard least squares approach be optimal from the perspective of Gauss-Markov theory. In situations where returns are heavy tailed and skewed alternative models will prove more appropriate as we will discuss.

Traditional approaches, both frequentist and Bayesian, to quantile regression have involved parametric models based on the asymmetric Laplace (AL) distribution. Using asymmetric Laplace distribution provides a mechanism for Bayesian inference of quantile regression models. Hu et al. (2012) develop a fully Bayesian approach for fitting single-index models in conditional quantile regression. The benefit of using a Bayesian procedure, lies in the adoption of available prior information and the provision of a complete predictive distribution for the required reserves (de Alba, 2002). Different Bayesian loss reserve models have been proposed for different types of claims data. Zhang {\em et al.} (2012) propose a Bayesian non linear hierarchical model with growth curves to model the loss development process, using data from individual companies forming various cohorts of claims. Ntzoufras and Dellaportas (2002) investigate various models for outstanding claims problems using a Bayesian approach via Markov chain Monte Carlo (MCMC) sampling strategy and show that the computational flexibility of a Bayesian approach facilitated the implementation of complex models.

\subsection{Contributions} \par \vspace{-4mm} \noindent
The contribution of this paper is three-fold. First, we propose using quantile regression for loss reserving. The proposed method, relating the provision to quantile regression allows a direct modelling of risk margin, and hence provision, instead of estimating the mean then applying a risk margin. It provides a richer characterization of the data, especially when the data is heavy tailed, allowing us to consider the impact of a covariate on the entire distribution, not merely its conditional mean. Secondly, we develop a range of parametric quantile regression models in Bayesian framework, each with their own distribution features. Especially, in particular we generalize the AL distribution model to incorporate a dynamic mean, variance and the shape parameters to model risk margin via a user friendly Bayeisan software {\tt WinBugs}, which is easy for users without much Bayeisan background or specialized knowledge of Markov chain Monte Carlo (MCMC) methodology.  Furthermore, the estimation of shape parameter by accident year gives us an analytical framework to estimate risk margin.  This allows us to capture the feature that the cohort of claims in different accident year may be heterogeneous, and hence applying different different risk margin to different accident year gives us an explicit provision in reserving. Finally, we compare the performance of parametric and nonparametric quantile regressions in the context of loss reserving.

The rest of the paper is organized as follows. Section 2 explains the parametric and non-parametric models proposed. Section 3 presents the posterior quantile regression models in a Bayesian framework. Section 4 details the way to calculate risk measures and risk margin using our models. Then we apply the methodology to two real loss reserve data sets in Section 5 and 6. Section 7 concludes.

\section{Quantile Regression for Claims Reserving} \par \vspace{-4mm} \noindent
In this section, we present quantile regression models and explain their relevance to loss reserving, this will be undertaken in both a non-parametric and a parametric modelling framework under the Bayesian paradigm. In the process we propose a novel analytical approach to perform estimation of the risk margin under various quantile regression model structures. Of particular focus in this paper is the class of models based on the Asymmetric Laplace (AL) distributional family. In the special case of the AL distribution we demonstrate that risk margin estimation is achieved naturally through the modelling the shape parameters of the AL distribution and hence the inference on the model parameters directly informs the inference of the risk margin.

In developing a quantile regression framework for general insurance claims development triangles we will assume that there is a run-off triangle containing claims development data in which $Y_{ij}$ will denote the cumulative claims with indices $i \in \left\{0, . . . , I\right\}$ and $j \in \left\{0, . . . , J\right\}$, where $i$ denotes the accident year and $j$ denotes the development year (cumulative claims can refer to payments, claims incurred, etc). Furthermore, without loss of generality, we make the simplifying assumption that the number of accident years is equal to the number of observed development years, that is, $I = J$ with $N=\frac{1}{2}I(I+1)$ observations. At time $I$ the index set in the {\em upper} triangular is
\begin{equation}
\mathcal{D}_o = \left\{(i,j): \; i + j \leq I+1 \right\}
\end{equation}
and for claims reserving at time $I$ the index set to predict the future claims in the {\em lower} triangle is:
\begin{equation}
\mathcal{D}_l = \left\{(i,j): \; i + j > I+1, \; i \leq I, \; j \leq I \right\}.
\end{equation}
Therefore the vector of observed $Y_{ij}$ in the upper triangle is given by $\bm Y_o = \left\{Y_{ij}: \, (i,j) \in \mathcal{D}_o \right\}$ and the corresponding vector of covariates is denoted by $\bm x_o= \left\{\bm x_{ij}: \, (i,j) \in \mathcal{D}_o \right\}$. Similarly $\bm Y_l= \left\{Y_{ij}: \, (i,j) \in \mathcal{D}_l \right\}$ and $\bm x_l= \left\{\bm x_{ij}: \, (i,j) \in \mathcal{D}_l \right\}$ are the vectors of claims and covariates in the lower triangle.

In the quantile regression structures we will aim to make inference on the quantile function of the data within sample, in each cell of $\bm Y_o$
as well as predictive out-off sample quantile function estimation based on the claim cells in $\bm Y_l$ in lower triangle.
The estimation of the quantile function regression has three main components:
\begin{itemize}
\item{The conditional distribution and in this case conditional quantile function of the dependent variables given by the claims data, given the explanatory variables.;}
\item{The structural component of the regression structure based on the link functions and imposed model structures linking the regression structures with the covariates to the location and scale of the conditional distribution and conditional quantile functions of the response.;}
\item{The actual choice of independent variables i.e. the covariates in the regression model, in this case we will also consider some basis function regression structures in some of the models proposed.}
\end{itemize}

In the following sub-sections we discuss each of these components in term, starting with the distributional aspects of the quantile regression models we consider.

\subsection{Nonparametric Quantile Regression Models} \par
\begin{figure}
\caption{\sf Loss function}
\hspace{40mm}
\includegraphics[height=5cm,width=9cm]{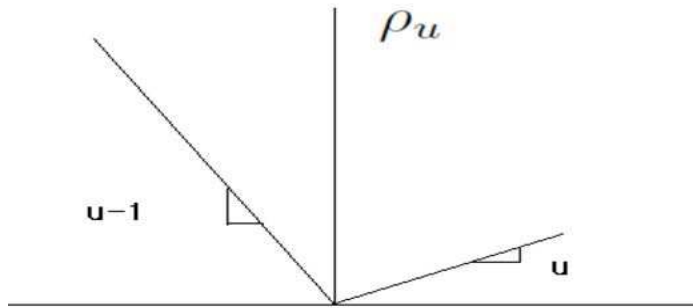} \par
\label{LossFcn}
\end{figure}

 \vspace{-4mm} \noindent
In a non-parametric quantile regression approach, we perform estimation of regression coefficients without the need to make any assumptions on the distribution of the response, or equivalently the residuals. If $Y_{ij}>0$ is a set of observed losses and $\bm x_{ij}=(1,x_{ij1},\dots, x_{ijm})$ is a vector of covariates that describe $Y_{ij}$. The quantile function for the log transformed data $Y_{ij}^*=\ln Y_{ij} \in \Re$ is
\begin{equation}
Q_{Y^*}(u|\bm x_{ij})=\alpha_{0,u}+\sum\limits_{k=1}^m \alpha_{k,u} \, x_{ijk} \label{NonPar}
\end{equation}
where
$u \in (0,1)$ is the quantile level, $\bm \alpha_u=(\alpha_{0,u},\dots, \alpha_{k,u})$ are the linear model coefficients for quantile level $u$ which are estimated by solving
\begin{equation} \label{EqnLossNonParQuant}
\min_{\alpha_{0,u}, \dots, \alpha_{m,u}} \sum_{i,j\le I} \rho_u(\epsilon_{ij})=\sum_{i,j \le I} \epsilon_{ij} [u-I(\epsilon_{ij}<0)]
\end{equation}
and $\epsilon_{ij}=y_{ij}^*-\alpha_{0,u}-\sum\limits_{k=1}^m \alpha_{k,u} \, x_{ijk}$. Then the quantile function for the original data is $Q_{Y}(u|\bm x_{ij})=\exp(Q_{Y^*}(u|\bm x_{ij}))$. Koenker and Hallock (2001) illustrate the loss function $\rho_u$ for quantile regression as we represent in Figure \ref{LossFcn}. \par \vspace{8mm}
\par \noindent
Koenker and Machado (1999) and Yu and Moyeed (2001) show that the solution to minimization of the loss function in equation (\ref{EqnLossNonParQuant}) for estimating the parameter vector $\bm \alpha_u$ is equivalent to maximum likelihood estimation of the parameters of the AL distribution. Hence, the parameter vector $\bm \alpha_u$ can be estimated via an AL distribution with pdf
\begin{equation} \label{PdfAL}
f(y_{ij}^*|\mu_{ij},\sigma_{ij}^2,p)=\frac{p(1-p)}{\sigma_{ij}}\exp\left( -\frac{(y^*_{ij}-\mu^*_{ij})}{\sigma_{ij}}[p-I(y_{ij}^* \le \mu_{ij})] \right)
\end{equation}
where the skew parameter $0<p<1$ gives the quantile level $u$, $\sigma_{ij}>0$ is the scale parameter and $-\infty<\mu^*_{ij}<\infty$ is the location parameter. Since the pdf (\ref{PdfAL}) contains the loss function (\ref{EqnLossNonParQuant}), it is clear that parameter estimates which maximize (\ref{PdfAL}) will minimize (\ref{EqnLossNonParQuant}).

In this formulation the AL distribution represents the conditional distribution of the observed dependent variables (responses) given the covariates. More precisely, the location parameter $\mu_{ij}$ of the AL distribution links the coefficient vector $\bm \alpha_u$ and associated independent variable covariates in the linear regression model to the location of the AL distribution. It is also worth noting that under this representation it is straightforward to extend the quantile regression model to allow for heteroscedasticity in the response which may vary as a function of the quantile level $u$ under study. To achieve this one can simply add a regression structure linked to the scale parameter $\sigma_{ij}$ in the same manner as was done for the location parameter.

Equivalently, we assume $Y_{ij}^*$ conditionally follows an AL distribution denoted by $Y_{ij}^*\sim AL(\mu^*_{ij}, \sigma_{ij}^2,u)$. Then
\begin{equation}
Y_{ij}^* =\mu^*_{ij}+\epsilon^*_{ij} \sigma_{ij}
\end{equation}
where $\epsilon^*_{ij} \sim AL(0,1,u)$, $\mu^*_{ij}=\alpha_{0,u}+\sum\limits_{k=1}^{m} \alpha_{k,u} \, x_{ijk}$ and $\sigma_{ij}^2=\exp(\beta_{0,u}+\sum\limits_{k=1}^{\nu} \beta_{k,u} \, s_{ijk})$.
Discussion on the choice of link function and structure of regression terms will be undertaken in later sections.
In presenting the model in this fashion we already start to move towards the representation of a parametric quantile regression structure.

\subsection{Parametric Quantile Regression Models} \par \vspace{-4mm} \noindent
Alternatively, we may adopt a parametric approach to study the quantile regression structure. Two types of distributions, on real support $\Re$ or positive support $\Re^+$ can be considered and we begin with distributions on $\Re$. In this case, we assume that $Y^*_{ij} \sim F(y^*|\bm \theta)$ where $F(y^*|\bm \theta)$ is the conditional cumulative distribution function (cdf) and $\bm \theta \in \bm \Theta$ is a vector of model parameters including all unknown coefficient parameters and distributional parameters. The quantile function for the conditional distribution of $Y^*_{ij}$ given $\bm x_{ij}$ at a quantile level $u\in(0,1)$ is given by:
\begin{equation}\label{quan}
Q_{Y^*}(u|\bm x_{ij}) \equiv \inf \left\{y^*: \; F(y^*|\bm \theta) \ge u \right\}.
\end{equation}
Under this formulation, the conditional quantile function in (\ref{quan}) can be written as
\begin{equation}
Q_{Y^*}(u|\bm x_{ij})= \mu^*_{ij}
+Q_{\epsilon^*}(u)\sigma_{ij} \label{QuanFcnPar}
\end{equation}
where $Q_{\epsilon^*}(u)=F_{z^*}^{-1}(u)$ is the inverse cdf for the standardized variable $Z_{ij}^*=\frac{Y_{ij}^*-\mu^*_{ij}}{\sigma_{ij}}$ and again one may incorporate regression structures given as follows for location and scale functions:
\begin{eqnarray}
\text{\textbf{location:}} \hspace{3mm} \mu^*_{ij} &= &\alpha_0+\sum\limits_{k=1}^{m} \alpha_k x_{ijk}, \label{EqnLoc} \\
\text{\textbf{scale:}} \hspace{7mm} \sigma_{ij}^2 &= & \exp(\beta_{0}+\sum\limits_{k=1}^{\nu} \beta_{k} s_{ijk}). \label{EqnScale}
\end{eqnarray}
To transform the quantile function $Q_{Y^*}(u|\bm x_{ij})$ back to the original scale of the data $Y_{ij}=\exp(Y_{ij}^*)$, we suggest $Q_{Y}(u|\bm x_{ij})=\exp(Q_{Y^*}(u|\bm x_{ij}))$. We note that there is no unique way to transform the quantile function $Q_{Y^*}(u|\bm x_{ij})$ for $Y^*_{ij}$ back to $Y_{ij}$ and the proposed transformation $Q_{Y}(u|\bm x_{ij})=\exp(Q_{Y^*}(u|\bm x_{ij}))$ does not equal in general to the quantile function for the log-AL distribution.

\noindent \textbf{Remark:} \textit{We observe that the difference between the non-parametric and the parametric quantile regression models is that in the parametric structure we make explicit the quantile function of the ``residual'' denoted by $Q_{\epsilon}(u)$.
}

For distributions on $\Re^+$, we assume that $Y_{ij} \sim F(y|\bm \theta)$ with mean $\exp(\mu^*_{ij})$ where $\mu^*_{ij}$ is given in (\ref{EqnLoc}). Next we make explicit several possible parametric models one may consider in quantile regressions for risk margin. Each model has different associated properties with regard to the relationship of the skewness, kurtosis and heaviness of the tail that it imposes on the quantile function of the response given the covariates.

\subsubsection{Asymmetric Laplace Distribution} \par \vspace{-4mm} \noindent
As discussed above, the AL distributional family is a useful model structure which naturally fits into a quantile regression framework. As made explicit above, the AL distribution is a three parameter distribution which has been shown to be directly linked to the estimation of quantiles in a quantile regression framework, see further details in Yu and Zhang (2005).

Since this realization, the AL family has been utilized in several financial risk and econometric settings such as Guermat and Harris (2001) who use the symmetric laplace distribution with GARCH volatility to model short-horizon asset returns. Lu et al. (2010) extend this to allow skewness via AL distribution. Yu and Moyeed (2001) apply AL distribution for quantile regression purposes, though as yet, no such developments have been made in the insurance and particularly the risk margin context. Here we propose such a model for risk margin estimation.

\begin{figure}[H]
\caption{\sf (a) The pdf of asymmetric Laplace distribution} \hspace{10mm}
\includegraphics[height=5cm,width=15cm]{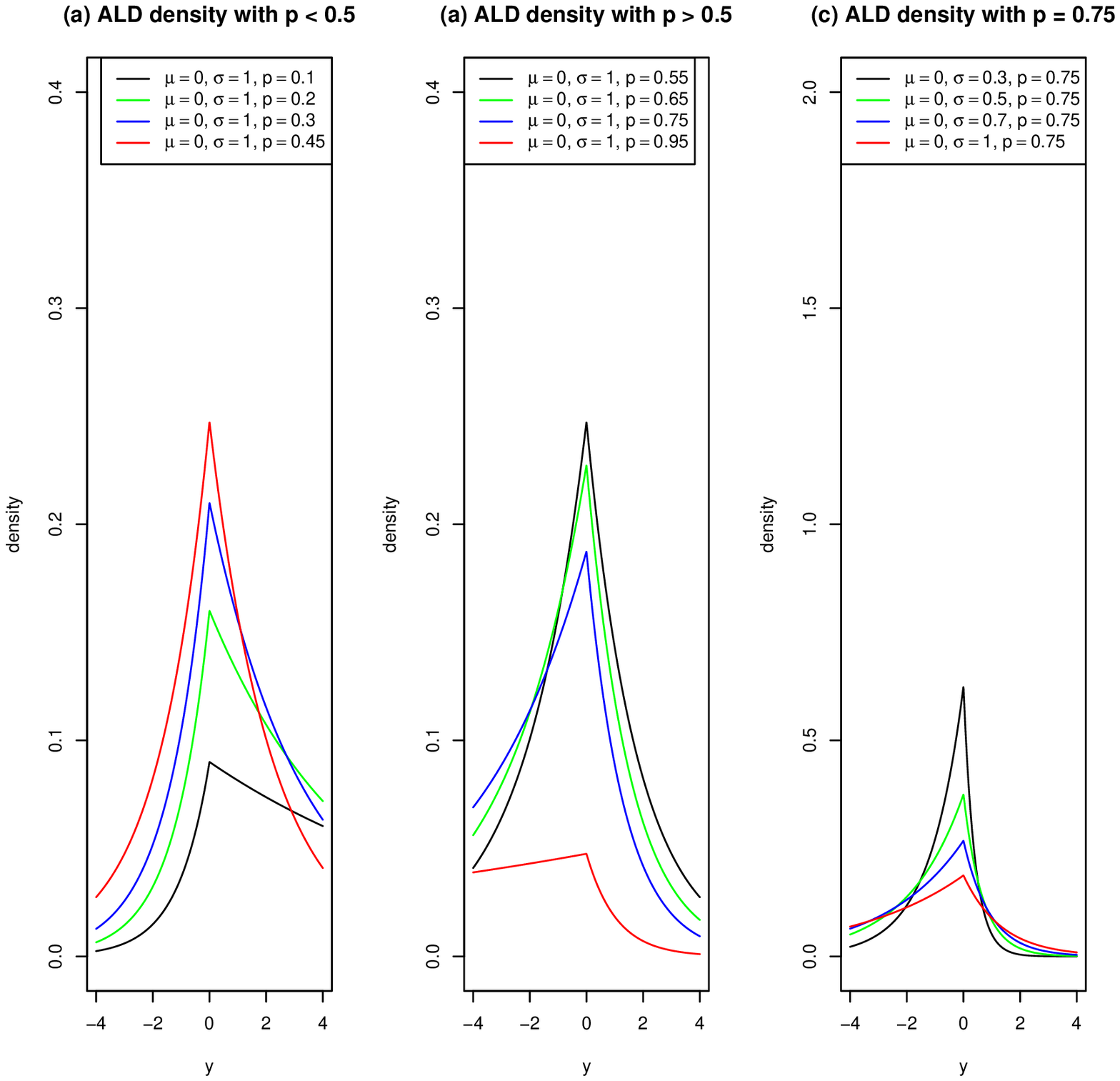} \par \noindent
{\hspace{20mm} Figure 2: \sf (b) The skewness and kurtosis of asymmetric Laplace distribution \hspace{20mm}} \par \hspace{30mm}
\includegraphics[height=6cm,width=10cm]{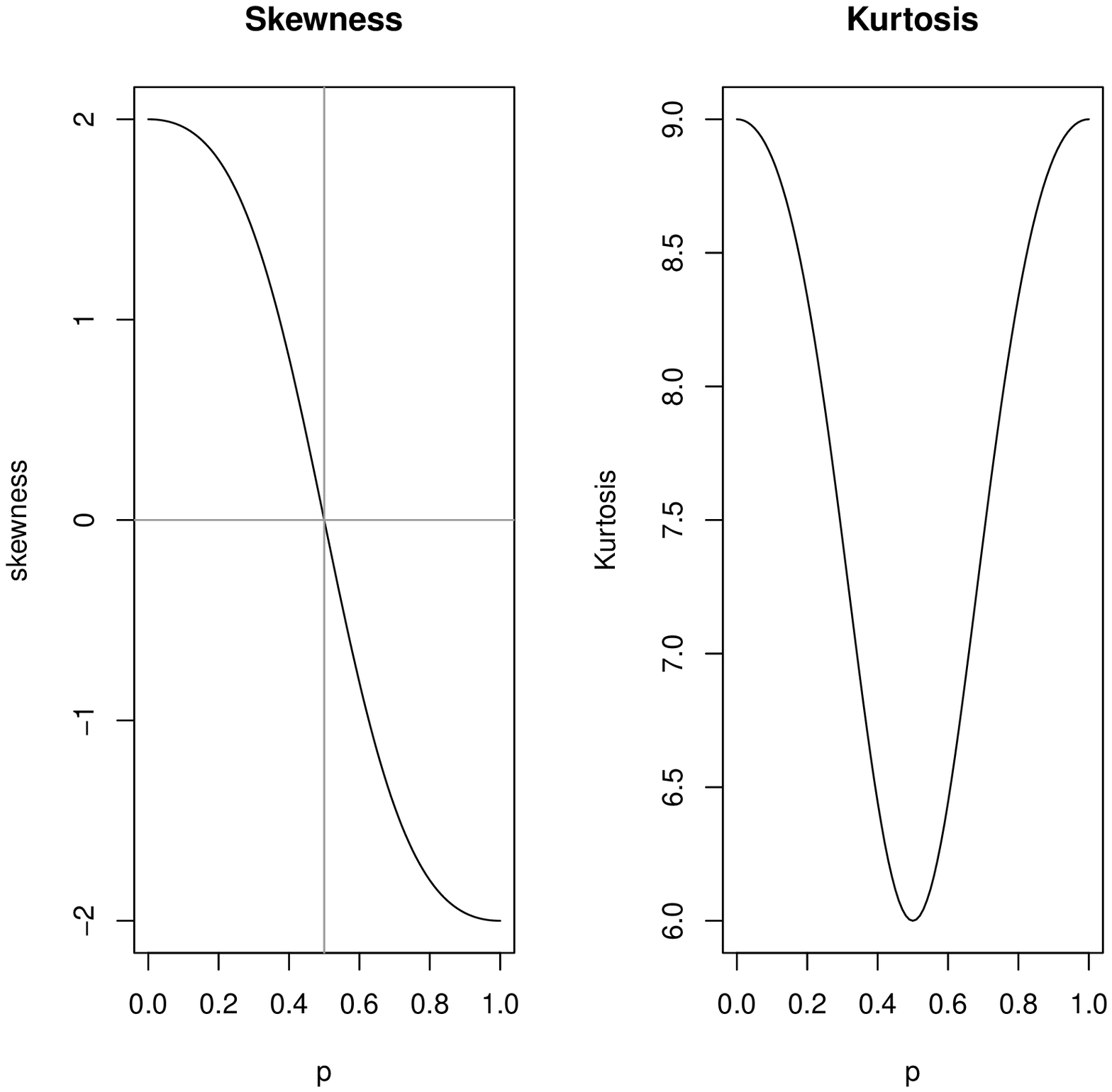} \par
\label{ALPdf}
\end{figure}

If we model the residuals $\epsilon_{ij}$ by an AL distribution, the quantile function for observed data $Y_{ij}^*$ is given by (\ref{QuanFcnPar}) where $F_{z^*}^{-1}(u)$ is the inverse cdf (quantile function)
\begin{equation}
 F_{AL}^{-1}(u|\mu,\sigma^2,p)= \left\{ \begin{array} {ll} \mu+\frac{\sigma}{1-p}\log(\frac{u}{p}), & \mbox{if} \ 0\leq u \leq p,\\ & \vspace{-4mm} \\
\mu-\frac{\sigma}{p}\log(\frac{1-u}{1-p}), & \mbox{if} \ p< u \leq 1. \end{array} \right. \label{InvCdfAL}
\end{equation}
\noindent To understand how the three location, shape and scale parameters of the AL distribution affect the shape and tails of the distribution it is also useful to note the following relationship between the parameters and the mean, variance, skewness $S$ and kurtosis $K$ of AL distribution:
\begin{eqnarray}
E(Y) & = & \mu+\frac{\sigma(1-2p)}{p(1-p)},\hspace{7mm} Var(Y) = \frac{\sigma^{2}(1-2p+2p^{2})}{(1-p)^{2}p^{2}}, \label{ALVar} \\
S(Y) & = & \frac{2[(1-p)^{3}-p^{3}]}{((1-p)^{2}+p^{2})^{3/2}}, \hspace{5mm} K(Y) = \frac{9p^{4}+6p^{2}(1-p)^{2}+9(1-p)^{4}}{(1-2p+2p^{2})^{2}}. \label{ALSkew}
\end{eqnarray}
Note that the shape parameter $p$ of the AL distribution gives the magnitude and direction of skewness. AL distribution is skewed to left when $p>0.5$ and skewed to right when $p<0.5$ and hence it can model the left skewness of most log transformed loss data directly through this shape parameter $p$. Moreover as the risk margin adopted in insurance industry is mostly greater than 50 percent, AL distribution allows the calculation of quantiles rather than  mean estimates fairly easily. Figures \ref{ALPdf}(a) and \ref{ALPdf}(b) show a variety of pdf for AL distribution and its skewness and kurtosis respectively.

 \subsubsection{Power Pareto Model} \par \vspace{-4mm} \noindent
As the second choice of parametric quantile regression model we consider the framework of Cai (2010). In this approach a polynomial power-Pareto (PP) quantile function model is developed. This model combines a power distribution with a Pareto distribution, which enables us to model both the main body and the tails of a distribution. In considering the PP model the conditional quantile function of the response (reserve in each cell) are comprised of two components:
\begin{itemize}
\item{component 1: a power distribution $F_1(y) = y^{\frac{1}{\gamma_1}}$ where $y \in [0,1]$ and $\gamma_1 > 0$ with a corresponding quantile function then given by $Q_1\left(u; \gamma_1\right) = u^{\gamma_1}$ for $u \in [0,1]$; and}
\item{component 2: a Pareto distribution function $F_2(y) = 1 - y^{-\frac{1}{\gamma_2}}$ where $y \geq 1$ and $\gamma_2 > 0$ with a corresponding quantile function then given by
$Q_2\left(u;\gamma_2\right) = \left(1-u\right)^{-\gamma_2}$.}
\end{itemize}
One may use the fact that the product of the two quantile functions will remain a strictly valid quantile function producing the new quantile function family known as the Polynomial-Power Pareto model. The resulting structural form given by the inverse cdf of the Pareto distribution with an additional polynomial power term:
\begin{equation}
 F^{-1}_{PP}(u|\gamma_1,\gamma_2) = u^{\gamma_1}(1-u)^{-\gamma_2}. \label{InvCdfPP}
\end{equation}
Hence the quantile function is again given by (\ref{QuanFcnPar}) where $Q_{\epsilon^*}(u)=F^{-1}_{PP}(u)$ and $Q_{Y}(u)=\exp(Q_{Y^*}(u))$.

From the specification of this quantile function, one may then derive the resulting pdf of the PP model for $Y_{ij}^*=\ln Y_{ij}$ which is given by
\begin{eqnarray*}
 f_{PP}(y_{ij}^*|\gamma_1,\gamma_2) & = & \frac{\displaystyle u_{ij}^{1-\gamma_1}(1-u_{ij})^{\gamma_2+1}}{\displaystyle \sigma_{ij} [\gamma_2u_{ij}+\gamma_1(1-u_{ij})]} \label{PPPdf}
\end{eqnarray*}
where $u_{ij}$ is given by solving the system of equations defined for each observation by
\begin{equation} \label{EqnPPCQF}
y_{ij}^* = \mu^*_{ij} + u_{ij}^{\gamma_1}\left(1-u_{ij}\right)^{-\gamma_2} \sigma_{ij}.
\end{equation}
where again we treat the location $\mu^*_{ij}=\mu^*_{ij}\left(\bm{\alpha}\right)$ in (\ref{EqnLoc}) and scale $\sigma_{ij}=\sigma_{ij}\left(\bm{\beta}\right)$ in (\ref{EqnScale}) as functions of the regression coefficients and associated covariates. We note that in this case the $u_{ij}$ is really an implicit function of the regression structure as each $u_{ij}$ is found as the solution to the system of equations in (\ref{EqnPPCQF}).

To complete the specification of the polynomial power Pareto model we plot the shape of the density that can be obtained for a range of different power parameters for the power and pareto components with a unit scale factor $\sigma = 1$. These plots in Figure \ref{PPPdf} demonstrate the flexible skewness, kurtosis and tail features that can be obtained from such a model by varying the parameters $\gamma_1$ and $\gamma_2$.
\begin{figure}[H]
\caption{\sf The pdf of Power Pareto distribution} \vspace{3mm} \hspace{25mm}
\includegraphics[height=6cm,width=10cm]{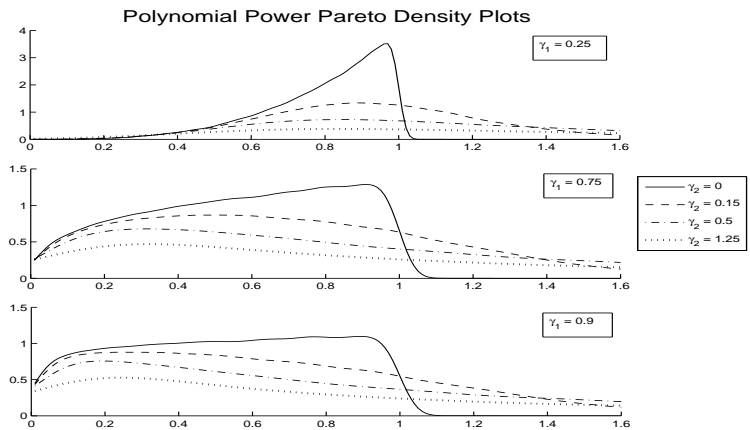}
\label{PPPdf}
\end{figure}

\subsubsection{Generalised Beta Distribution of the Second Type Family} \par \vspace{-4mm} \noindent
We note that the AL and PP families of quantile regression models require a log transformation of the data before the modelling to ensure the data has real support $\Re$ that these distributions are defined upon. In performing this transformation, one must analyze carefully the effect of the transformation on the ability to fit such models and the resulting model interpretability must be interpreted with regard to the transformation. This is particularly the case if zero counts are present in the data for some accident and development years. Moreover, in the context of claims reserving, loss data often exhibits heavy-tailed behavior, particularly for long tail business classes. To account for such features and to remove the need to consider pre-transformation of the data one may consider the family of generalized beta (GB2) distributions of the second kind.

The type two generalized beta distribution (GB2) has attractive features for modelling loss reserve data, as it has a positive support $\Re^+$ and nests a number of important distributions as its special cases. The GB2 distribution has four parameters, which allows it to be expressed in various flexible densities. See Dong and Chan (2013) for a more detailed description of GB2 distribution including its pdf and distribution family.

If $Y_{ij} \in \Re^+$ conditionally follows a GB2 distribution, then it can be characterized by the density given by
\begin{equation}
f_{GB2}(y_{ij}|a,b_{ij},p,q) = \frac{\frac{a}{b_{ij}}(\frac{y_{ij}}{b_{ij}})^{ap-1}}{B(p,q)[1 + (\frac{y_{ij}}{b_{ij}})^{a}]^{p+q}},
\hspace{3mm}\mbox{for $y_{ij}\geq0$} \label{GB2Pdf}
\end{equation}
where $a,p$ and $q$ are shape parameters and $b_{ij}$ is the scale parameter.

In particular, $b_{ij}$ can be linked to the mean $\mu_{ij}$ of the distribution as follows:
\begin{equation}
b_{ij}= \frac{\mu_{ij}B(p,q)} {B(p+1/a,q-1/a)}
\end{equation}
where $\mu_{ij}$ is log-linked to a linear function of covariates $\mu^*_{ij}$ in (\ref{EqnLoc}) according to the relationship:
\begin{equation}
E(Y_{ij}) =\mu_{ij}=\exp\left(\alpha_0+\sum_{k=1}^m \alpha_k \, x_{ijk} \right). \label{mfcn}
\end{equation}
Then the variance is given by:
\begin{equation}
Var(Y_{ij})= \mu^2_{ij} \left\{\frac{B(p,q)B(p+2/a,q-2/a)} {[B(p+1/a,q-1/a)]^2} -1\right\}.  \label{GB2var}
\end{equation}
The GB2 distribution is a generalization from the beta distribution with pdf:
\begin{eqnarray}
f_{B}(z_{ij}|p,q)=\frac{1}{B(p,q)}z_{ij}^{p-1}(1-z_{ij})^{p+q}
\end{eqnarray}
via the transformation $\displaystyle z_{ij}=\frac{(\frac{y_{ij}}{b_{ij}})^a}{1+(\frac{y_{ij}}{b_{ij}})^a}$.
Hence the cdf of GB2 distribution is given by:
\begin{equation}
F_{GB2}(y_{ij}|a,b_{ij},p,q) = \int_{0}^{z_{ij}} \frac{t^{p-1}(1-t)^{(q-1)}}{B(p,q)} dt=\frac{B(z_{ij}|p,q)}{B(p,q)} = F_B(z_{ij}| p,q) \label{F}
\end{equation}
where $B(z_{ij}|p,q)$ is the incomplete beta function.

The GB2 is directly relevant for quantile regression models since one may also find its quantile function in closed form according to the following expression:
\begin{equation}
Q_Y(u)=\frac{\exp \left(\alpha_0+\sum\limits_{k=1}^m \alpha_k x_{ijk} \right)B(p,q)} {B(p+1/a,q-1/a)} \left(\frac{F_B^{-1}(u| p,q)}{1-F_B^{-1}(u| p,q)}\right)^{\frac{1}{a}}. \label{QuanFcnGB2}
\end{equation}

There are many widely known and utilized sub-families of the GB2 family, we present two examples of relevance to the context of risk margin estimation that we will explore, corresponding to the generalized gamma and the gamma distribution sub-families.

\subsubsection{Two Special Cases of GB2} \par \vspace{-4mm} \noindent
To understand the flexibility of the GB2 family of models, we consider the case when $q=\infty$, then the resulting GB2 distribution sub-family becomes the generalized gamma (GG) distribution, see discussion in McDonald et al. (1984). The GG family of models was independently introduced by Stacy (1962), as a three parameter distribution with pdf given by:
\begin{equation}
f_{GG}(y_{ij}|a,b_{ij},p) = \lim\limits_{q\rightarrow \infty} \frac{\frac{a}{b_{ij}}(\frac{y_{ij}}{b_{ij}})^{ap-1}}{B(p,q)[1 + (\frac{y_{ij}}{b_{ij}})^{a}]^{p+q}}=\frac{a(\frac{y_{ij}}{b_{ij}})^{ap}\exp[-(\frac{y_{ij}}{b_{ij}})^a]}{y_{ij}\Gamma(p)},
\hspace{3mm}\mbox{for $y_{ij}>0$} \label{TB}
\end{equation}
where $a$ and $p$ are shape parameters and $b_{ij}$ is scale parameter linked to the mean of the distribution as:
\begin{equation}
b_{ij}= \frac{\mu_{ij}\Gamma(p)} {\Gamma(p+1/a)} \label{EyGG1}
\end{equation}
and the mean is again log-linked to a linear function of covariates in (\ref{mfcn}). The cdf is
$$ F_{GG}(y_{ij}|a,b_{ij},p) = \int_{0}^{z_{ij}} \frac{t^{p-1}e^{-t}}{\Gamma(p)} dt=\frac{\gamma_1(z_{ij}|p)}{
\Gamma(p)} = F_G(z_{ij}| 1,p)$$
where $\gamma_1(z_{ij}|p)$ is the lower incomplete gamma function and $z_{ij}=(\frac{y_{ij}}{b_{ij}})^a$. Hence, the quantile function is given by:
\begin{equation}
Q_Y(u)=\frac{\exp \left(\alpha_0+\sum\limits_{k=1}^m \alpha_k x_{ijk} \right)\Gamma(p)} {\Gamma(p+1/a)} \left(F_G^{-1}(u| 1,p)\right)^{\frac{1}{a}} \label{TBq}
\end{equation}
\par \vspace{3mm} \noindent
The second case is nested within the GG family and corresponds to the two parameter Gamma distribution which is obtained by further restricting $a=1$. Its pdf and quantile function are well-known and can be expressed using equations (\ref {TB}) and (\ref{TBq}) by replacing $a$ with 1.

Having defined clearly the three different quantile regression distributional families that will be considered in the parametric quantile regression framework, we now introduce the different regression structures we consider in the quantile regression under each distributional assumption.

\subsection{Structural Components of the Quantile Regression Framework} \par \vspace{-4mm} \noindent
In the model structures we will adopt, as is standard practice in regression modelling, once we believe we have suitable explanatory variables for the dependent variable quantity of interest, in this case the conditional quantile function, we will assume the observations are independent.

In the following subsections we explain how under each different distributional assumption for the conditional quantile regression structure, one may introduce a link function to relate regression models using independent covariates to the response quantiles in order to model trend behaviors in the location and scale of the quantile function. To simplify all the possible different model considerations we consider only log link functions in all regressions.

The possible regression structures we consider will be classified as: location based explanatory factors i.e. trends in accident and development years; and scale (heteroskedascity / variance) based explanatory factors for accident and development years. We note that when it comes to different distributional choices since we may transform the observations, we are actually considering both additive and multiplicative (mixed interaction) terms in our regressions and as such we explore aspects of ANOVA as well as ANCOVA regression structures in the quantile regression setting. A summary of the model structures we consider for the location and scale components of each model is provided in Table \ref{TabModelStruct} in Appendix 1. We note that in general one may consider that a version of the ANCOVA model was applied to the PP and AL models and a version of the ANOVA model was effectively applied to the AL and GB2 families. In addition we will allow the influence of covariates to affect different quantile levels to different extents, making for an interesting analysis on the effect of model structure on quantile level.

We note that since the focus of this manuscript differs to that undertaken in the Poisson-Tweedie regression context of Peters, Shevchenko and Wuethrich (2009), in that the focus of the regression model comparison will be primarily concerned with the model choice for the distributional form of the conditional quantile function, not so much on the model structure uncertainty related to all possible covariate model sub-space structures and nested models, therefore we limit the analysis to the ANOVA and ANCOVA structures given below. If one is interested in specialized techniques to explore and compare all possible models sub-spaces within each distributional model, we suggest the approach adopted in Peters, Shevchenko and Wuethrich (2009) or recently in Verrall, R.J. and Wuthrich, M. (2013)

\subsubsection{Location: Development and Accident Year Trend Model Structures} \par \vspace{-4mm} \noindent
The primary sets of covariates we consider correspond to the accident year and the development year in the claims reserving structure, as well as transformations of these through basis functions. From Table \ref{TabModelStruct} one may observe that we label models using two subscripts according to their mean and variance functions respectively.  Models 0$\stackrel{\tiny\bullet}{}$ (denoted by $M_{0 \cdot}$) and 1$\stackrel{\tiny\bullet}{}$ (denoted by $M_{1 \cdot}$) are parsimonious location structure specifications for the general model in (\ref{EqnLoc}) with $m=2$, that is, the additive structure is given by:
\begin{eqnarray}
 \mbox{Model 0}\hspace{-1mm}\stackrel{\tiny\bullet}{}: \hspace{3mm} \mu^*_{ij} & = & \alpha_{0} + \alpha_{1} \times i + \alpha_{2} \times j, \label{u0} \\
 \mbox{Model 1}\hspace{-1mm}\stackrel{\tiny\bullet}{}: \hspace{3mm} \mu^*_{ij} & = & \alpha_0 + \alpha_1^S F_1(j) + \alpha_2^C F_2(j).
\end{eqnarray}
Under $M_{0 \cdot}$ one assumes a linear trend across accident and development years. If a non-linear trend across development years is considered with an assumption of common behavior down the accident years, on may consider $M_{1 \cdot}$ which is a basis regression model popular in term structure models and known as the Nelson-Seigel model (Nelson and Siegel, 1987). Examples of typical basis functions we considered under this choice for the location are given in Figure \ref{FigNSBasis} below, where we show the `level', `slope' and `curvature' structure of the location trend from such a model.

\begin{figure}
\caption{\sf Basis function regression structure for development years in location parameter in the AL model ($M_{1\cdot}$). Decomposition of the role the level, slope and curvature basis functions play in the regression with example coefficients: $\alpha_0 = 1$, $\alpha_1^S = 0.5$, $\alpha_2^C = 2$ and $\lambda = 0.5$ with $j \in \left\{1,2\ldots, J\right\}$ in years.}
\includegraphics[height=6cm,width=\textwidth]{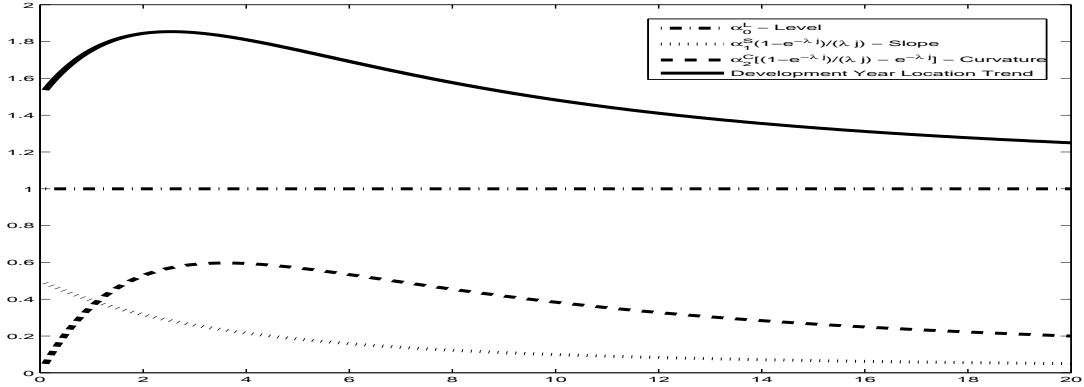}
\label{FigNSBasis}
\end{figure}

In the context of an ANOVA model specification for the location one can assume a form given by:
\begin{eqnarray}
 \mbox{Model 2}\hspace{-1mm}\stackrel{\tiny\bullet}{}: \hspace{3mm} \mu^*_{ij} & = & \alpha_{0} + \alpha_{1i} + \alpha_{2j}. \label{u1}
\end{eqnarray}
This location (trend) function corresponds to the general model in (\ref{EqnLoc}) with $m=2$,
$$ \alpha_1 x_{ij1}=\alpha_{1i} \hspace{3mm} \mbox{and} \hspace{3mm} \alpha_2 x_{ij2}={\alpha}_{2j}.$$
The parameters $\alpha_{1i}$ and ${\alpha}_{2j}$ denote the accident year and development year effects respectively and they satisfy the following constraints:
\begin{eqnarray}
\alpha_{11}  =  {\alpha}_{21} = 0.
\end{eqnarray}
This parametrization is set up in the context of loss reserving so that all parameters are relative to the first accident year which has the most information. These location functions (\ref{u0}) to (\ref{u1}) apply to both AL and PP distributions in general. For Gamma, GG and GB2 distributions with positive support $\Re^+$, a log link function is considered and the location functions become $\mu_{ij}=\exp(\mu_{ij}^*)$. When the AL distribution, with the shape parameter $p=u$ is applied, Model 3$\stackrel{\tiny\bullet}{}$ ($M_{3 \cdot}$) corresponds a nonparametric quantile function
\begin{eqnarray}
 \mbox{Model 3}\hspace{-1mm}\stackrel{\tiny\bullet}{}: \hspace{3mm} \mu^*_{ij,u} & = & \alpha_{0,u} + \alpha_{1i,u} + \alpha_{2j,u} \label{u2}
\end{eqnarray}
where $\alpha_{_{\bullet},u}$ are parameters at quantile level $u$.

\subsubsection{Scale: Development and Accident Year Variance Model Structures} \par \vspace{-4mm} \noindent

There are different choices for the structure of the variance function for the AL and PP distributions but Gamma, GG and GB2 distributions do not have a component to model $\sigma^2$ directly. Model $\stackrel{\tiny\bullet}{}$\,0 ($M_{\cdot 0}$) assumes homoscedastic variance $\sigma^2_{ij}=\sigma^2$. Models $\stackrel{\tiny\bullet}{}$0 ($M_{\cdot 0}$) to $\stackrel{\tiny\bullet}{}$3 ($M_{\cdot 3}$)
are specified below:
\begin{eqnarray}
 \mbox{Model} \ \stackrel{\tiny\bullet}{}\hspace{-1mm}0: \hspace{3mm} \sigma^{2}_{ij} & = & \sigma^2 , \label{u2} \\
 \mbox{Model} \ \stackrel{\tiny\bullet}{}\hspace{-1mm}1: \hspace{3mm} \sigma^{2}_{ij} & = & \exp(\beta_0 + \beta_{1i}) , \label{u2} \\
 \mbox{Model} \ \stackrel{\tiny\bullet}{}\hspace{-1mm}2: \hspace{3mm} \sigma^{2}_{ij} & = & \exp(\beta_0 + {\beta}_{2j}), \label{u3} \\
 \mbox{Model} \ \stackrel{\tiny\bullet}{}\hspace{-1mm}3: \hspace{3mm} \sigma^{2}_{ij} & = & \exp(\beta_0 + \beta_{1i} + {\beta}_{2j}) \label{Varij},
\end{eqnarray}
where the parameters $\beta_{1i}$ and ${\beta}_{2j}$ which denote the accident year and development year effects respectively satisfy the following constraints:
\begin{eqnarray}
\beta_{11}  =  {\beta}_{21} = 0.
\end{eqnarray}
Again Models $\stackrel{\tiny\bullet}{}$1 to $\stackrel{\tiny\bullet}{}$2 corresponds to (\ref{EqnScale}) with $\beta_1 s_{ij1}=\beta_{1i}$ and $\beta_2 s_{ij2}=\beta_{2j}$.
Furthermore, for Model 23', the shape parameter in the AL distribution is further modelled by the accident year effect, which is specified as follows:
\begin{eqnarray}
 \mbox{Model \ 23'}: \hspace{3mm} p_{i} & = & \phi_0 + \phi_{1i} . \label{u5}
\end{eqnarray}
where the parameters $\phi_{1i}$ denote the accident year effect and satisfy the following constraints:
\begin{eqnarray} \phi_{11} = 0. \end{eqnarray}

\section{Bayesian Framework: Posterior Quantile Regression} \par \vspace{-4mm} \noindent
The estimation of quantile regression models is straightforward to adopt under a Bayesian formulation. One of the key advantage of using Bayesian procedures for practical models such as those we develop above lies in the adoption of available prior information and the provision of a complete predictive distribution for the required reserves (de Alba, 2002).

To complete the posterior distribution specification in each model it suffices to consider the representation of two components: the likelihood of the data for the regression structure (that is, the density not the quantile function); and the prior specifications for the model parameters. In the above sections, the quantile function of the likelihood is presented, along with the associated density for the observations conditional upon the parameters and covariates, that is, the likelihood for each model. Therefore, to formulate the Bayesian structure we simply need to present the prior structures we consider for the parameters in each model. This will be relatively straightforward for models formed from the AL distribution structure and the GB2 structures, but not so trivial for the case of the PP model.

In the real data examples we consider below, we adopt an objective Bayesian perspective in which we consider relatively uninformative priors. This reflects our lack of prior knowledge for the model parameters likely ranges or magnitudes. For instance, the priors for parameters (coefficients) in mean, variance and skewness quantile regression functions are all selected as Gaussian:
\begin{equation}
\alpha_{0}, \ \alpha_{1},  \ \alpha_{1}^S,  \ \alpha_{1i}, \ \alpha_{2},\ \alpha_{2}^C, \ {\alpha}_{2j}, \ \beta_{1i}, \ {\beta}_{2j}, \ \phi_0, \ \phi_{1i} \sim N(0,100)
\label{p1}
\end{equation}
and for the shape parameters of the GB2 distribution are:
\begin{equation}
a \sim N(0,100), \hspace{7mm} p \sim Ga(0.001,0.001), \hspace{7mm} q \sim Ga(0.001,0.001). \label{p2}
\end{equation}
Normal and gamma distributions are standard choices of priors for parameters with a real and positive support respectively, see discussions on possible choices in Denison et. al. (2002). In the case of the AL and GB2 models, these priors combined with the resulting likelihoods produce in each case standard and well defined posterior distributions.

In the case of the PP model one has to be careful to define the posterior support to ensure the resulting distribution is normalized and therefore a proper posterior density. To ensure this is the case one must impose constraints on the posterior support which can be uniquely characterized by the three sets of parameter space constraints $\Omega_1$, $\Omega_2$ and $\Omega_3$, for coefficient vectors $\bm{\alpha}$, $\bm{\beta}$ and $\left(\gamma_1,\gamma_2\right)$ respectively, given by:
\begin{equation}
\begin{split}
&\Omega_1 = \left\{\left(\alpha_{0,u},\ldots,\alpha_{m,u}\right)\; : \; \alpha_{0,u} + \sum_{k=1}^m \alpha_{k,u}x_{ijk} < y_{ij}, \;\; \forall i,j \in \left\{1,2,\ldots,I\right\} \right\},\\
&\Omega_2 = \left\{\left(\beta_{0,u},\ldots,\beta_{\nu,u}\right)\; : \; \beta_{0,u} + \sum_{k=1}^{\nu} \beta_{k,u}s_{ijk} > \epsilon > 0, \;\; \forall i,j \in \left\{1,2,\ldots,I\right\} \right\},\\
&\Omega_3 = (0,M]\times (0,\infty), \;\; M \in \Re^+.\\
\end{split}
\end{equation}
Under these parameter space restrictions the resulting posterior for the PP model can be shown to be well defined as a proper density, see a derivation and proof in Theorem 1 of Cai (2010).

In Cai (2010) they consider an MCMC scheme for the resulting posteriors based on standard Metropolis-Hastings steps with rejection when the proposed parameter values fail to satisfy the posterior support constraints. In general this results in a very slowly mixing MCMC chain which will have very poor properties. We replace this idea with simple block Metropolis within Gibbs updates which allow for smaller moves in each component of the constrained posterior support making it more likely to satisfy the constraints and also simpler to design and tune the proposal for the MCMC scheme. This was a significant improvement compared to the approach proposed in Cai (2010). We implement this sampler in {\tt R}.
For the other Bayesian models from the AL and GB2 models, sampling from the intractable posterior distributions involved the Gibbs sampling algorithm (Smith and Roberts, 1993; Gilks
et al., 1996) and Metropolis-Hastings algorithm (Hastings, 1970; Metropolis et al., 1953) are the most popular MCMC techniques. For readers who are less familiar with Bayesian computation techniques, we suggest the {\tt WinBUGS} (Bayesian analysis Using Gibbs Sampling) package, see Spiegelhalter et al. (2004). The MCMC algorithms that are implemented for each model in {\tt WinBugs} and {\tt R} are available upon request.

In the Gibbs sampling scheme, a single Markov chain is run for 60,000 to 110,000 iterations, discarding the initial 10,000 iterations as the burn-in period to ensure convergence of parameter estimates. Convergence is also carefully checked by the history and autocorrelation function (ACF) plots. The every 10-th simulated values from the Gibbs sampler after the burn-in period are sampled to mimic a random sample of size 5000 to 10,000 from the joint posterior distribution for posterior inference. Parameter estimates are given by the posterior means.

\section{Quantile Prediction for Risk Measures, Risk Margin} \par \vspace{-4mm} \noindent
\label{SectionQuantilePred}
As discussed in the introduction, the predicted reserves are typically performed in a claims reserving setting by predicting the mean reserve in each cell in the lower triangle $\mathcal{D}_l$. Other methods for reserving may involve the quantification of a risk measure based on the distribution of the predicted reserves, in place of the mean predicted reserve, such as VaR, Expected Shortfall (ES) or Spectral Risk Measures (SRM), see discussions in the tutorial review of Peters et.al. (2013). In addition, in order to quantify the uncertainty in a central measure for the predicted reserve, one may alternatively take the central measure of reserve and make a risk margin adjustment based on the distribution of the predicted reserves in the form of a quantile function.

When calculating any of these required measures for the resulting total outstanding reserves one requires to first obtain the predictive density, which under the Bayesian setting can be obtained for instance in one of the following two ways for each $Y_{ij} \in \mathcal{D}_l$:
\begin{itemize}
\item{\textbf{Full Predictive Posterior Distribution:}
\begin{equation*}
F_{Y_{ij}}\left(y_{ij}|\mathcal{D}_0\right)
= \int_{0}^{y_{ij}} f_{Y_{ij}}\left(y | \mathcal{D}_0 \right)\; dy = \int_{0}^{y_{ij}} \int f_{Y_{ij}}\left(y | \bm{\theta}\right) \pi\left(\bm{\theta} | \mathcal{D}_0 \right)\; d \bm{\theta}\; dy  .
\end{equation*}
Here, all posterior parameter uncertainty is integrated out of the predictive distribution.
}
\item{\textbf{Conditional Predictive Posterior Distribution:}
\begin{equation*}
F_{Y_{ij}}\left(y_{ij}|\widehat{\bm{\theta}}\left(\mathcal{D}_0\right)\right) = \int_{0}^{y_{ij}} f_{Y_{ij}}\left(y | \widehat{\bm{\theta}}\left(\mathcal{D}_0\right)\right) \; dy
\end{equation*}
where the point estimator $\widehat{\bm{\theta}}\left(\mathcal{D}_0\right)$ contains the information from the upper triangle. Examples of common estimators include the posterior mean $\widehat{\bm{\theta}}\left(\mathcal{D}_0\right) = \widehat{\bm{\theta}}^{(MMSE)}$ or mode $\widehat{\bm{\theta}}\left(\mathcal{D}_0\right) = \widehat{\bm{\theta}}^{(MAP)}$.
}
\end{itemize}
Using these predictive distributions, one may also be interested in quantities such as the distribution of the total outstanding claim given by the sum of the losses in the lower triangle according to the random variable $Y_T:= \sum\limits_{(i,j) \in \mathcal{D}_l} Y_{ij}$ which has distribution given under the full predictive posterior distribution according to convolution given by  
\begin{equation} \label{EqnConvExp}
\begin{split}
F_{Y_T}\left(y_t|\widehat{\bm{\theta}}\left(\mathcal{D}_0\right)\right)&:=\ast_{(i,j) \in \mathcal{D}_l} F_{Y_{ij}}\left(y|\widehat{\bm{\theta}}\left(\mathcal{D}_0\right)\right)\\
&= \left(F_{Y_{I,2}} \ast F_{Y_{I-1,3}} \ast F_{Y_{I-2,4}} \ast \cdots \ast F_{Y_{I,I}}\right)\left(y|\widehat{\bm{\theta}}\left(\mathcal{D}_0\right)\right).
\end{split}
\end{equation}
where, one convolves the distributions for the loss elements in the lower triangle with $\ast$ the convolution operator. One can then state several features about the tail behavior of the total loss distribution and also therefore of the high quantiles as $y \rightarrow \infty$, depending on the properties of the individual loss random variables in the sum. For instance, if one has loss distributions on $\Re^+$ then one can obtain the lower bound given by
\begin{equation}
\begin{split}
\overline{F_{Y_T}}\left(y_t|\widehat{\bm{\theta}}\left(\mathcal{D}_0\right)\right) &:= \overline{
\left(F_{Y_{I,2}} \ast F_{Y_{I-1,3}} \ast F_{Y_{I-2,4}} \ast \cdots \ast F_{Y_{I,I}}\right)}\left(y|\widehat{\bm{\theta}}\left(\mathcal{D}_0\right)\right)\\
&\sim c \sum_{(i,j) \in \mathcal{D}_l} \overline{F_{Y_{ij}}}\left(y|\widehat{\bm{\theta}}\left(\mathcal{D}_0\right)\right), \;\; \text{as} \; y \rightarrow \infty,
\end{split}
\end{equation}
for some $c \geq 1$. Note, if at least one of the lower triangle losses $Y_{ij}$ is distributed according to a heavy tailed loss distribution, such as sub-exponential, regularly varying or long tailed loss distributions then one can find the precise value for $c$. For instance if the total loss is max-sum equivalent, then $c=1$, see definitions for regular variation, sub-exponential, long tailed and max-sum equivalence in Bingham et al. (1989) and in the context of insurance and quantile function approximations as discussed here, see the recent tutorial and references therein from Peters et al. (2013).

These conditional predictive distributions can be obtained for any model approximately by solving the integrals using the Markov chain Monte Carlo samples obtained from the posterior $\pi\left(\bm{\theta} | \mathcal{D}_0 \right)$. Then, given a predictive distribution, one can then find quantile functions according to the following approaches:
\begin{itemize}
\item{\textbf{Full Predictive Posterior Quantile Function:} is given by
$Q_{Y_{ij}|\mathcal{D}_0}\left(u\right):=F^{-1}_{Y_{ij}}\left(y_{ij}|\mathcal{D}_0\right)$ which is the solution to the second order ordinary differential equation:
\begin{equation*}
\frac{d}{dQ_{Y_{ij}|\mathcal{D}_0}}f_{Y_{ij}}\left( Q_{Y_{ij}|\mathcal{D}_0}\left(u\right) |\mathcal{D}_0\right)\left(\frac{dQ_{Y_{ij}|\mathcal{D}_0}}{du}\right)^2 + f_{Y_{ij}}\left( Q_{Y_{ij}|\mathcal{D}_0}\left(u\right) |\mathcal{D}_0\right)\frac{d^2Q_{Y_{ij}|\mathcal{D}_0}}{du^2} = 0,
\end{equation*}
which is obtained by twice differentiating the following identity:
\begin{equation}
F_{Y_{ij}}\left(Q_{Y_{ij}|\mathcal{D}_0}\left(u\right)|\mathcal{D}_0\right) = \int_{0}^{Q_{Y_{ij}|\mathcal{D}_0}\left(u\right)} f_{Y_{ij}}\left(y | \mathcal{D}_0 \right) dy = u.
\end{equation}
The solution to this second order ordinary differential equation can often be found in the form of a power series, see discussions in Gyorgy and Shaw (2008).
}
\item{\textbf{Conditional Predictive Posterior Quantile Function:}
\begin{equation} \label{Eqn:pred2}
Q_{Y_{ij}|\widehat{\bm{\theta}}\left(\mathcal{D}_0\right)}\left(u\right):= F^{-1}_{Y_{ij}}\left(u|\widehat{\bm{\theta}}\left(\mathcal{D}_0\right)\right)
\end{equation}
which is the most convenient choice that we recommend since the inverse of the predictive distribution in this case takes the closed form expressions for the particular model considered as detailed in Section 2.2.
}
\item \textbf{Conditional Total Reserve Posterior Quantile Function:} In many cases one is also interested in finding the quantile function of the distribution corresponding to the total reserve, which under conditional independence is given by $F^{-1}_{Y_T}\left(y_t|\widehat{\bm{\theta}}\left(\mathcal{D}_0\right)\right)$ where this is given by the quantile function of the distribution in Equation \ref{EqnConvExp}. In general finding the convolution and inverse of this convolved distribution must be done numerically. There are many basic results known about these quantities such as asymptotic results and bounds for different properties of light and heavy tailed random variables, independent or dependent, see a discussion in Kaas et al. (2000).

\textbf{Light Tailed Run-off for Claims Process:} In the case in which no loss cells in the claims triangle are heavy tailed, then in general one would need to approximate the tail quantile for the partial sum of all losses. In Kaas et al. (2000) they study partial sums of random variables with  no assumption of independence or of identical marginal distributions. The only assumption is that the tails are not so heavy for each marginal, such that each marginal has finite mean. It will be useful to recall that for two random variables $X$ and $Y$, $X$ proceeds $Y$ under convex ordering $X \leq_{CX} Y$ iff for all convex real functions $g(\cdot)$ with finite expectations one has
\begin{equation}
\mathbb{E}\left[g(X)\right] \leq \mathbb{E}\left[g(Y)\right].
\end{equation}
Thus, two random variables X and Y with equal mean are convex ordered if their cdfs cross once.

Then one can show that in such cases for any sequence of loss distributions $\left\{F_{Y_{ij}}\right\}_{(i,j) \in \mathcal{D}_l}$ the following convex order relationship holds
\begin{equation}
\sum\limits_{(i,j) \in \mathcal{D}_l} Y_{ij} \leq_{CX} \sum\limits_{(i,j) \in \mathcal{D}_l} F_{Y_{ij}}^{-1}(U)
\end{equation}
for $U \sim U[0,1]$, see derivations in Goovaerts et al. (2000). This result means that the total loss $Y_T$ in the convex order sense, comprised of the most risky joint vector of losses with given marginals, has the comonotonous joint distribution. The components of which are maximally dependent since all components are non-decreasing functions of a common random variable $U$.

Hence, we consider the following quantile function approximation for the total loss based on the most conservative estimate using the above bound, given by
\begin{equation}
F^{-1}_{Y_T}(u) = \sum\limits_{(i,j) \in \mathcal{D}_l} F_{Y_{ij}}^{-1}(u). \label{QuantileForTotal}
\end{equation}
Note, in the case of heavy tailed losses this can be refined for large quantiles as follows.

\textbf{Heavy Tailed Run-off For Claims Process:} Alternatively, if additional features of the loss distributions in the lower triangle are known, such as these loss models contain at least one heavy tailed loss distribution, then one can bound the total quantile function result. This can be done conservatively by instead considering the $\mathcal{T}$-fold convolution of the distribution, say $F_{Y_{i*j*}}^{(*\mathcal{T})}$ which correspond to the loss distribution amongst all the lower trianglular loss elements with the dominant index of regular variation (that is, with the heaviest tails). In such cases it would be popular to utilize an asymptotic result for the quantile function of the sum, as the quantile level becomes large $u \d \rightarrow 1$. For instance, one could use the first order or second order asymptotic results, see discussions in Peters et al. (2013) and Cruz et al. (2014). As an example, if the quantile regression was structured such that the distribution of the partial sum $Y_T = \sum_{(i,j) \in \mathcal{D}_l} Y_{ij} \sim F_{Y_T}$ is regularly varying with index $ \rho  \geq 0$ with conditionally i.i.d. $Y_{ij}$ with each $Y_{ij}$ taking positive support, then one can write the first order tail approximation which is asymptotically equivalent to the following
\begin{equation}
\overline{F}_{Y_T}(y) \sim \mathcal{T} \overline{F}_{Y_{i*j*}}(y), \;\; y \rightarrow \infty,
\end{equation}
see detailed tutorial in Peters et al. (2013). This would lead to the approximation of the required quantile asymptotically by the expression
\begin{equation}
\begin{split}
Q_{Y_{T}|\widehat{\bm{\theta}}\left(\mathcal{D}_0\right)}\left(u\right)&:= \inf\left\{y\in \mathbb{R}^+ :\; F_{Y_T}(y) > u\right\}\\
%
%
%
&\approx \inf \left\{y\in \mathbb{R}^+ :\; \mathcal{T} \overline{F}_{Y_{i*j*}}(y) < 1-u\right\}\\
%
&\approx Q_{Y_{i*j*}|\widehat{\bm{\theta}}\left(\mathcal{D}_0\right)}\left(1-\frac{1-u}{\mathcal{T}}\right):= F^{-1}_{Y_{i*j*}}\left(1-\frac{1-u}{\mathcal{T}}| \widehat{\bm{\theta}}\left(\mathcal{D}_0\right)\right)
\end{split}
\end{equation}

\end{itemize}

\section{Model Structure Analysis for Israel data} \par \vspace{-4mm} \noindent
In this section we perform two core studies: The first involves isolating the structural components for the quantile regressions, in order to perform a study on the mean function and variance functions that are most suitable for an example of a representative claims reserving data set. This is therefore performed using the non-parametric and Bayesian formulations of the AL model with different assumptions on the mean and variance functions. The second involves isolating the distributional choices of the quantile regression, where we take the best fitting parametric model mean and variance function structures and use these to study distributional properties under the different quantile function choices.

The data set used throughout this section is interesting for such a benchmark exercise as it has been previously studied and its features are reasonably well known, see Chan et al. (2008) for more details on the Israel Data set. The data is available in Figure \ref{IsraelData} in Appendix I and represents the paid out claim amounts $y_{ij}$ for an Israel insurance company, covering periods from 1978 to 1995, containing 171 observations. For mathematical convenience, two zero claim amounts have been replaced with 0.01. Some general trends are observed in this data. Given an accident year, the claim development amounts generally increase between the first 4 to 6 development years then this increase is followed by a generally decreasing trend thereafter. The mean, median, variance and kurtosis of this data are 4459.7, 3,871, 12,059,232.6 and -0.4 respectively. The overall skewness is  0.58 and on a log scale is -2.67.

This data has been studied in Chan et al. (2008) using the generalized-$t$ (GT) distribution expressed as scale mixtures of uniform which facilitates the Bayesian implementation. They adopt the ANOVA and ANCOVA mean structures to study the accident year and development year effects on the conditional mean functions but not on any quantile level. Moreover they also remark that the log transformed data become negatively skewed which the symmetric GT distribution fails to accommodate. Hence, they suggest to adopt some skewed error distributions to improve the model performance.

Our primary point of departure for these previous studies on this data is the conjecture that using a measure of average effects may not be appropriate for understanding loss reserves at higher quantiles. Higher quantile projection is critical in loss reserving, for reinsurance premium calculations and also in deriving the risk margin. In this section, we use all the models in Section 2 for quantile projection with an aim to provide a more comprehensive study on model performance with a wide range of distributions having different tails behavior and model structures for the quantile trends and heteroscedasticity in the accident and development years.

\subsection{Analysis of Quantile Regression Models: Location and Scale} \par \vspace{-4mm} \noindent
To investigate the model structures for location (mean) and scale (variance) functions, we consider two settings: the first class of models involves the parametric models using the AL distribution with $p$ either fixed (denoted by fix) or left to be estimated (denoted by est), the mean functions given by (\ref{u0}) to (\ref{u1}) and variance being constant (Models 00-20) or given by (\ref{Varij}) (Models 03-23); the second class of models involves a set of nonparametric models which are also studied with mean function (\ref{u1}) and variance being constant or given by (\ref{Varij}) (Models 30 and 33) using AL as a proxy distribution with $p$ fixed at different quantile levels.

For model comparison, deviance information criterion (DIC) is adopted, see Appendix III for details. Since, models with smaller $DIC$ are preferred to those with larger $DIC$, then the results of the model comparisons provided in Table \ref{NPParModelFit} show that among the parametric models, $M_{23}$ which incorporates an ANOVA model for both accident and development years in modelling both the mean and variance functions is the best fitting model according to $DIC$. This show that the accident year and development year effects are both important in describing the dynamics of the mean and variance. Hence, these ANOVA-type mean and variance functions are applied to most of the subsequent analyses whenever possible. For the nonparametric models, $M_{33}$ with ANOVA variance provide better fit than $M_{30}$ with constant variance. \par
\begin{figure}[H]
\caption{\sf QQ plot for nonparametric models $M_{33}$ at different quantile levels} \hspace{10mm}
\includegraphics[height=10cm,width=14cm]{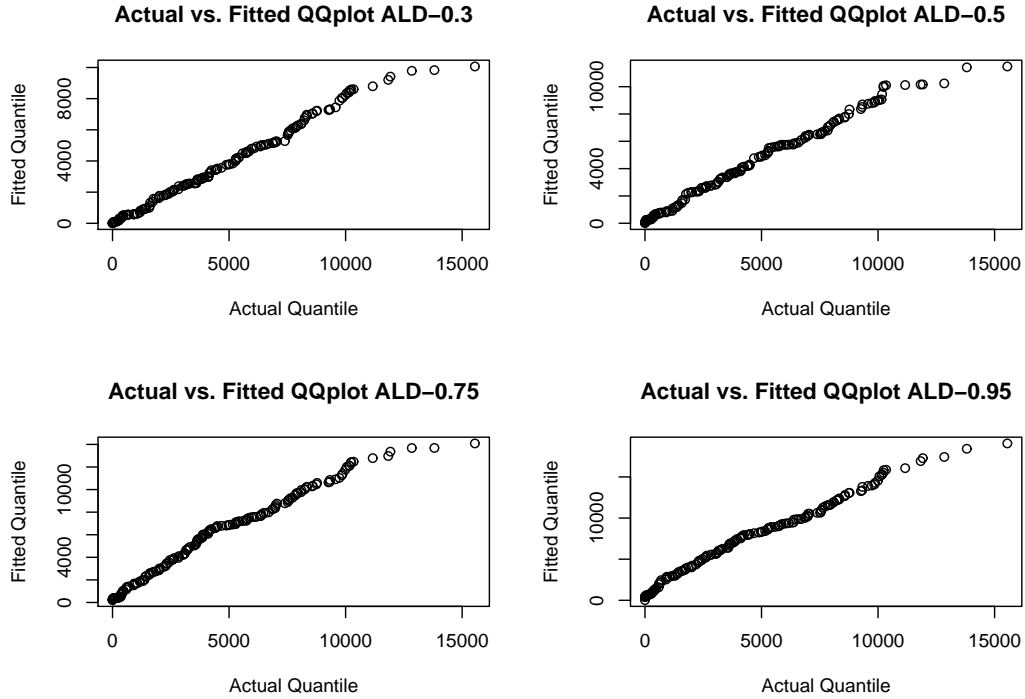} \par
\label{QQPlot}
\end{figure}

 \vspace{-2mm}
\begin{table}[H]
\caption{Estimates of $p$ and model fit measures for AL parametric and non-parametric models} \par \vspace{2mm}
\tabcolsep 5pt
\begin{tabular} {|l|c|c|c|c|| l|c|c|c|c|} \hline \label{NPParModelFit}
Models	 &	~~$DIC$~~	&	~~$\bar{D}^{\dag}$~~ &	~~$\hat{D}^{\ddag}$~~ & $p$	&	Models	& ~~$DIC$~~	&	 ~~$\bar{D}^{\dag}$~~ &	 ~~$\hat{D}^{\ddag}$~~	&	~~$p$~~ \\ \cline{2-10}
& \multicolumn{4} {|c||} {Variance Constant} & & \multicolumn{4} {c|} {Variance Function} \\ \hline
$M_{00}$ &195.41	 &255.21	 &315.02 &0.85 (est) & $M_{03}$ &272.82 &334.74	 &396.66   &0.93 (est)\\
$M_{10}$ &223.30  &284.10	 &344.91 &0.88 (est) & $M_{13}$ & 199.14 &247.49	 &295.85   &0.95 (est)\\
$M_{20}$ &50.94	 &120.17	 &189.40 &0.81 (est) & $M_{23}$ & -20.81 &24.91	 &70.63    &0.75 (est)\\ \hline
$M_{30}$ &55.94	 &125.61	 &195.28 &0.30 (fix) & $M_{33}$ & -37.06 &38.34	 &113.74  & 0.30 (fix)\\
$M_{30}$ &73.10   &152.26	 &231.43 &0.50 (fix) & $M_{33}$ & -38.80 &35.51	 &109.82  & 0.50 (fix)\\
$M_{30}$ &55.26	 &132.56	 &209.87 &0.75 (fix) & $M_{33}$ & -17.33 &53.40	 &124.12  & 0.75 (fix) \\
$M_{30}$ &44.86	 &116.38	 &187.91 &0.95 (fix) & $M_{33}$ & -64.26 &3.68	 &71.62  & 0.95 (fix)\\ \hline
\multicolumn{10} {l} {\footnotesize $\dag$~$\bar{D}$ is the posterior mean deviance $E_{\theta}[-2 \log f(\bm y| \bm \theta)]$; $\ddag$~$\hat D=-2 \log f(\bm y| \bar{\bm \theta})$ where $\bar{\bm \theta}$ is the posterior mean of $\bm \theta$}
\end{tabular}
\end{table}
Between parametric model $M_{23}$ and nonparametric models $M_{33}$, the nonparametric models provide better model performance according to $DIC$.
These models correspond to the AL models with mean and variance functions and we study their performances for a range of fixed quantile levels $p \in \left\{0.3,0.5,0.75,0.95 \right\}$ as shown in Figure \ref{QQPlot}. This plot demonstrates the quantile-quantile plot for the fitted models at different quantile levels, indicating appropriate fits from the specified model structures for a range of different quantile levels.

\begin{figure}[H]
\caption{\sf Fitted loss of the first accident year across quantiles using $M_{33}$ with AL distribution} \hspace{20mm}
\includegraphics[height=5.5cm,width=12cm]{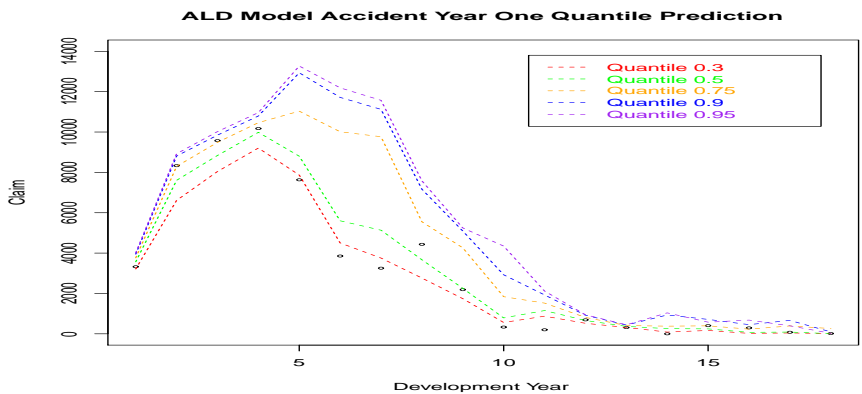}
\label{YearOneQuan}
\end{figure}

In addition, we investigate the trends of development year effects as depicted in Figure \ref{YearOneQuan} which reports the fitted loss $\widehat Y_{1j}=\exp(\mu_{1j}^*)$ where $\mu_{1j}^*$ is given by (\ref{u1}) and calculated using the conditional predictive posterior quantile function in (\ref{Eqn:pred2}) for the first accident year ($i=1$). The quantile levels $u$ correspond to the shape parameter $p$ set to 0.3, 0.5, 0.75 and 0.95 respectively in AL distribution. The figure demonstrates that there is a clear requirement for a nonlinear trend in the development year covariate at all quantile levels which uniformly increases up until $j=4$ and subsequently decreases thereafter at all quantile levels. Furthermore, the trends of fitted loss at all quantile levels agree with this observed trend.

 \par \vspace{-5mm}

\begin{figure}[H]
\caption{\sf Fitted loss of the upper triangle across quantiles using $M_{33}$ with AL distribution} \par \vspace{2mm} \hspace{5mm}
\includegraphics[height=4.5cm,width=8cm]{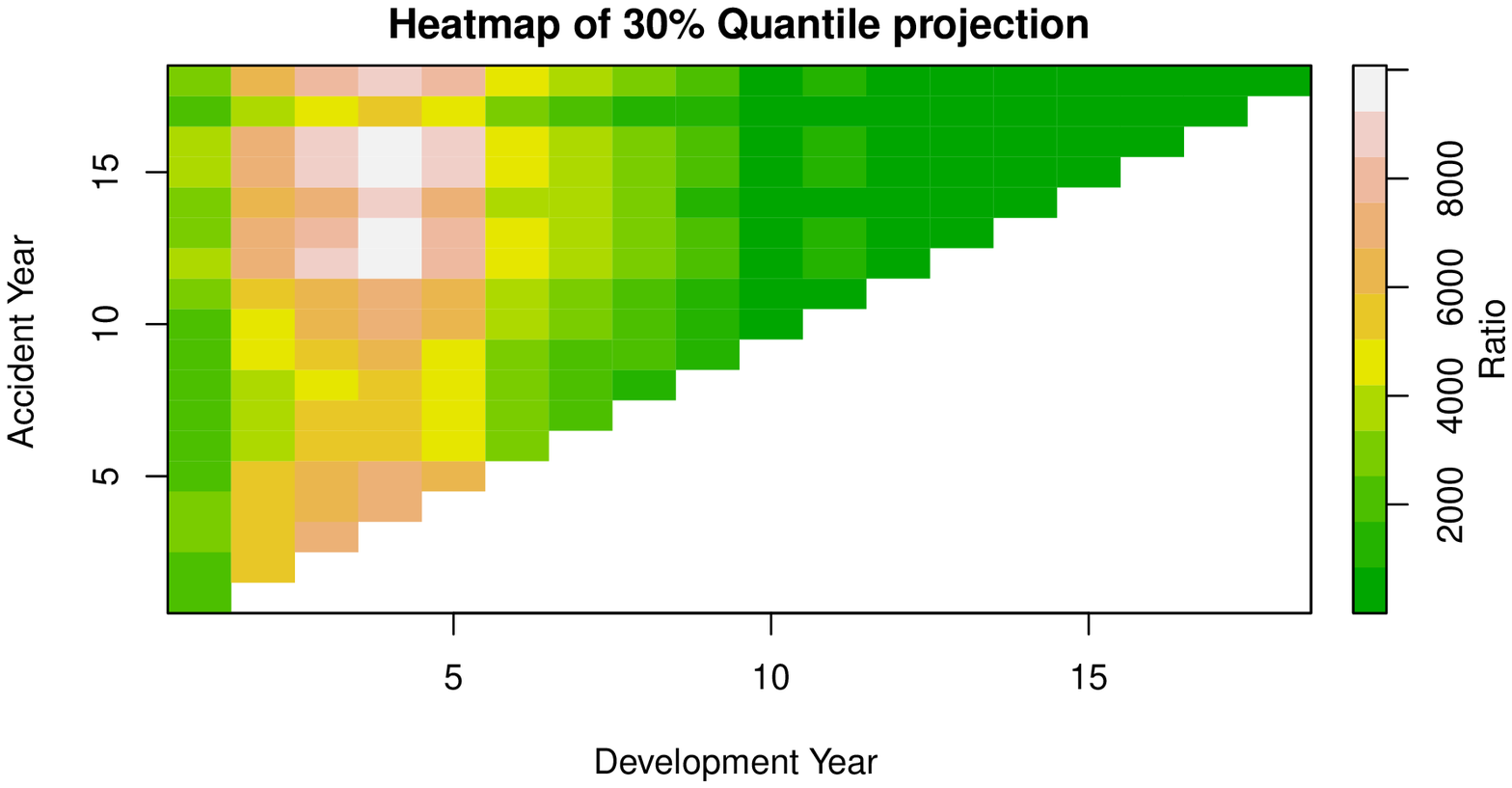} \includegraphics[height=4.5cm,width=8cm]{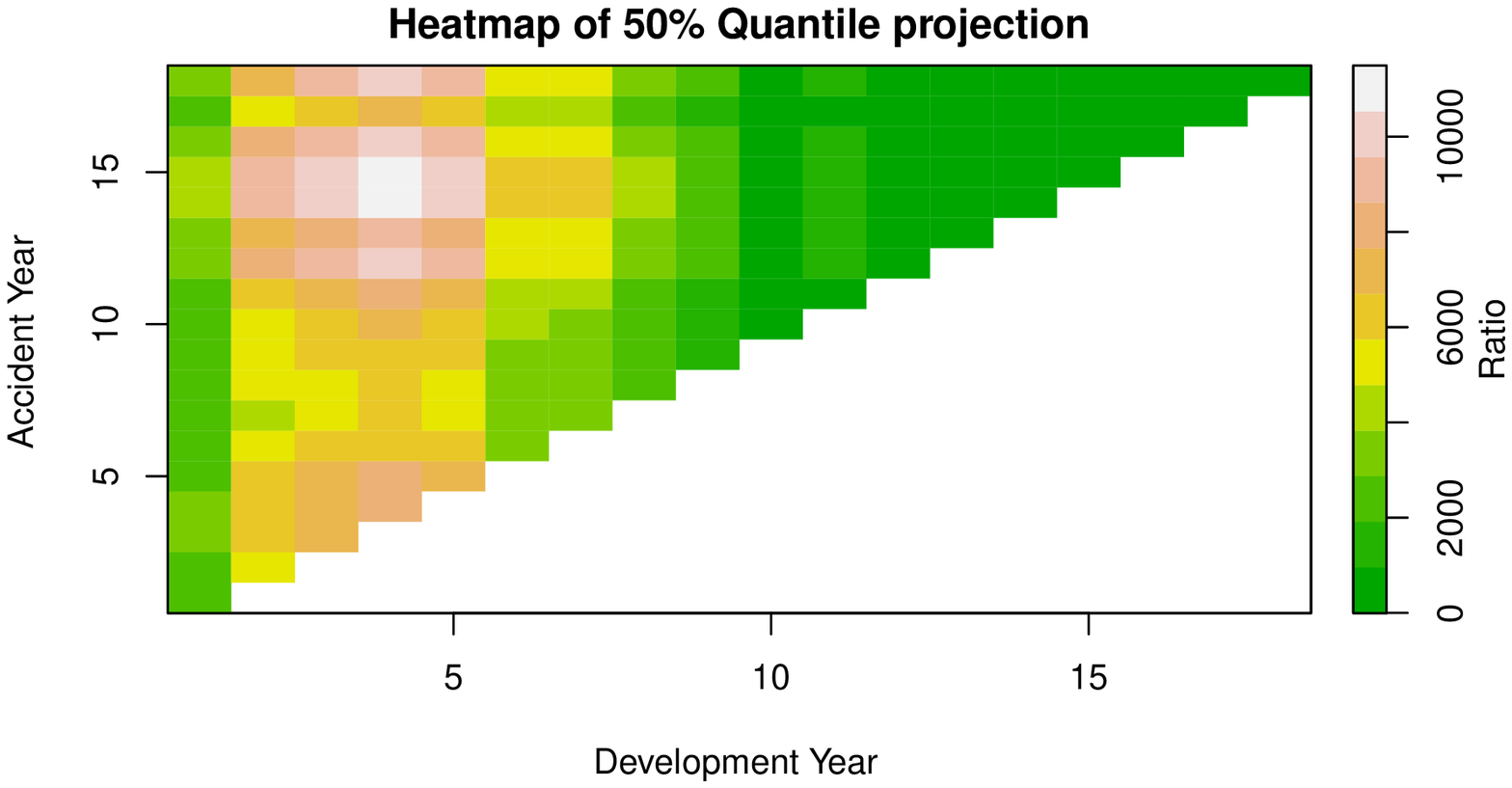} \par \hspace{5mm}
\includegraphics[height=4.5cm,width=8cm]{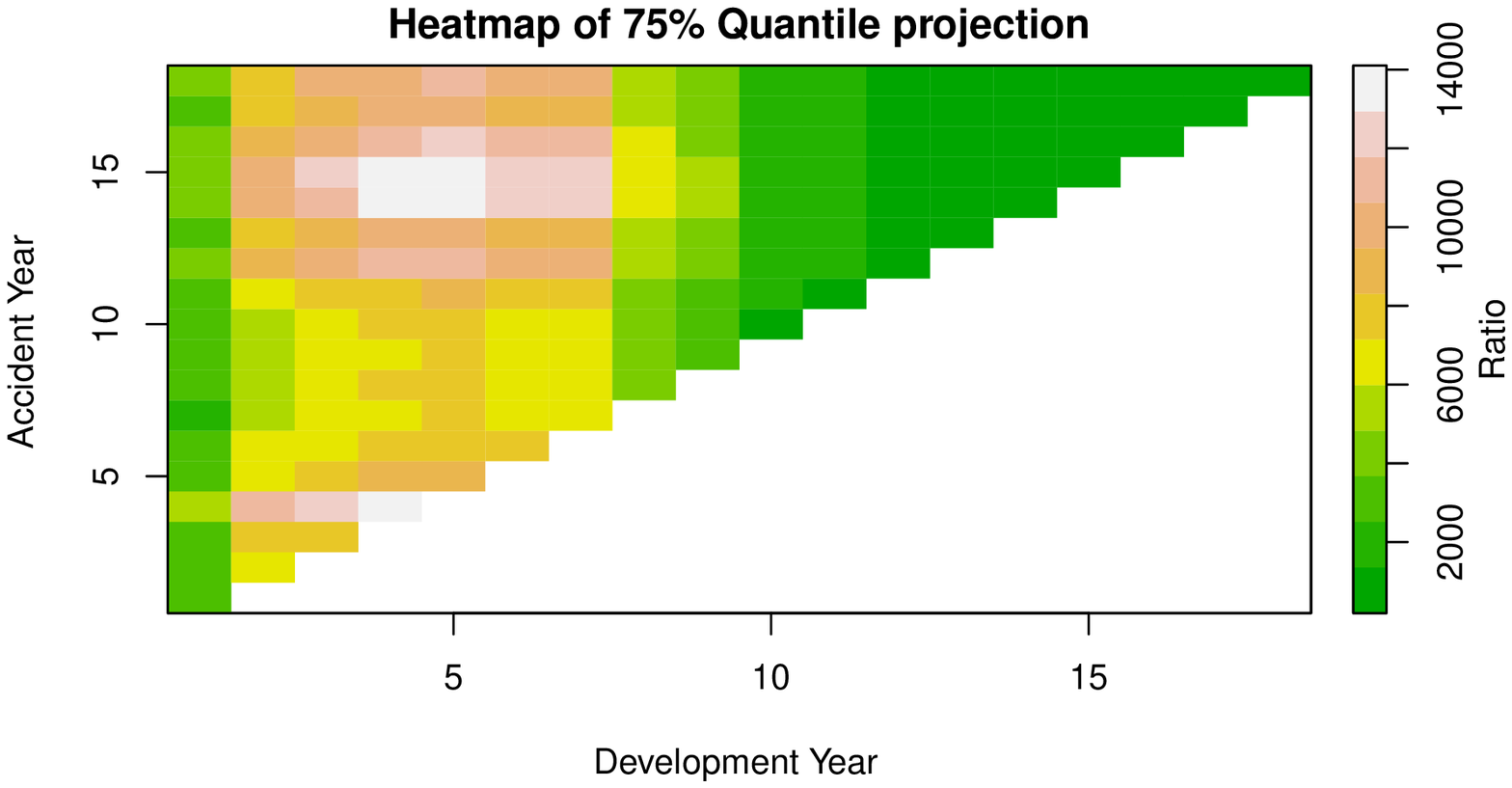} \includegraphics[height=4.5cm,width=8cm]{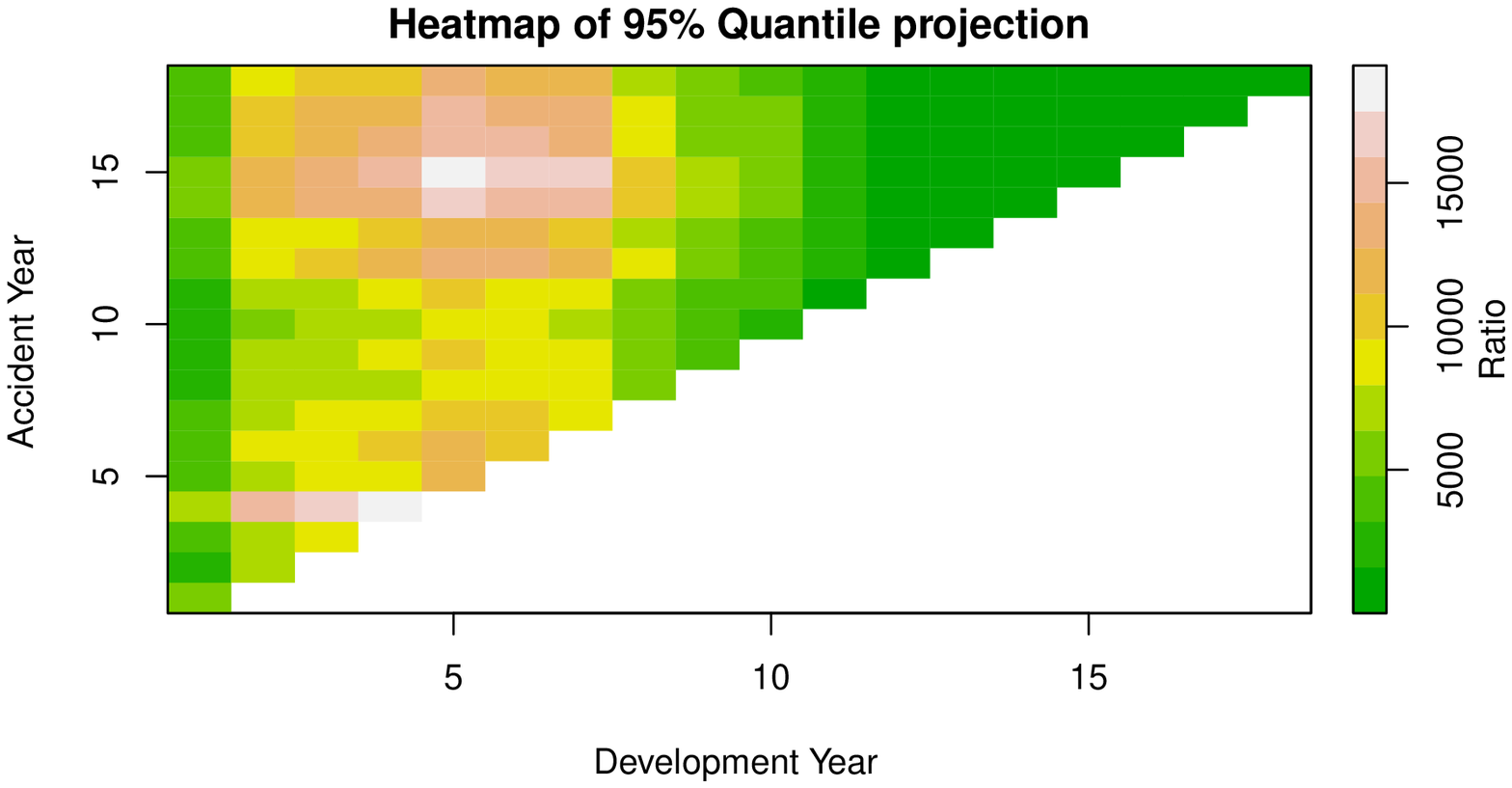} \par \vspace{-3mm} \hspace{5mm}
\label{HeatMap}
\end{figure}

To conclude the benchmark analysis on model structure we also present for the best model $M_{33}$ with mean and variance functions the estimated model trends for all accident years, depicted in Figure \ref{HeatMap} as five triangular heat maps. The heat maps each depict the fitted loss by accident and development years in the upper triangle at all five quantile levels, where the first row corresponds to that which was studied in Figure \ref{YearOneQuan}.
All heat maps show a consistent trend across development years for all accident years and quantile levels with high levels of loss as indicated by light colours being around the fourth development year, particularly for lower accident years. With increasing quantile levels, the width of light colours for each accident year increases showing higher levels of fitted losses around the peak.

Although nonparametric models have lower $DIC$ values, Table \ref{NPParModelFit} shows that parametric model $M_{23}$ actually provides comparable model fit according to $\bar D$s before model complexity penalty was applied. This is because parametric models with additional shape parameters are subject to heavier model complexity penalty. However it should be noted that parametric models provide better model fit in general over a range of models and quantile levels. In addition, the parametric models have a significant advantage that they will be more readily interpretable as well as directly usable when calculating risk margins and quantile based risk measures as long as the quantile functions are in closed form, as was discussed in Section \ref{SectionQuantilePred}. For the mean structure corresponding to model choice $M_{2\cdot}$ under parametric model we also studied different variance structures, in order to explore the different choices of variance functions under the AL distribution.

\begin{table}[H]
\caption{Parameter estimates and model fit measures for AL models with ANOVA mean and various variance functions} \par \vspace{2mm} \hspace{30mm}
\begin{tabular} {|l|r|r|r|r|c|c|c|} \hline	\label{ParModelFit}
Models	 &	~$DIC$~	&	~~~$\bar{D}$~~~ &	~~~$\hat{D}$~~~	&	~$MSE$~	 &	 ~$p$~ & ~ $\sigma^2$~  \\\hline
$M_{20}$ &  50.94	&120.17 &189.40 &	1015.71 &	0.80 &	0.02 	 \\
$M_{21}$ &  -4.32	&56.66 &117.64 &	849.91	 &	0.74 &	0.04 	 \\
$M_{22}$ &  6.63	&54.29 &101.95 &	755.66	 &	0.68 &	0.19 	 \\
$M_{23}$ &  -20.81	& 24.91&70.63 &	850.10 &	0.75 &	0.17 	 \\ \hline
\end{tabular}
\end{table}

Again, we confirm that amongst all models with AL distribution, $M_{23}$ which incorporates both accident and development year effects for the mean and variance demonstrates the best model fit according to $DIC$. On the other hand, $MSE$ favors $M_{22}$ which adopts only development year effect for the variance. One possible reason might be that the payments made in different accident years are relatively stable compared to those across development years, and hence the development year effect dominates in the variance estimation.

\subsection{Analysis of Quantile Regression Models: Quantile Distribution} \par \vspace{-4mm} \noindent

In this section we analyze the different model choices from the distributional perspective. This is not directly trivial to achieve, since each model has different features that must be taken into consideration in the comparison. It is clear from previous studies that one should always utilize an ANOVA-type mean function with accident and development years effect ($M_{2\cdot}$), or at a minimum incorporate a quadratic or basis function form for the development year effects such as $M_{1\cdot}$. In the case of the GB2 and AL models we will therefore consider mean structures in  $M_{2\cdot}$. However, in the case of the PP model we will consider $M_{1\cdot}$, since purely from a computational perspective it will be easier to implement an efficient MCMC sampler for $M_{1\cdot}$ compared to $M_{2\cdot}$. The reason for this is due to the rejection stage in the Metropolis-Hastings acceptance probability where under the PP model the posterior constraint regions will be easier to satisfy with less model complexity. In terms of the variance functions, when working with the GB2 models, we will consider $M_{2\cdot}$ in which we do not specify variance functions as there is no variance parameter in the distribution to model the variance directly. The variance of
the models are given by (\ref{GB2var}). Then in the case of the AL model we consider $M_{20}$ as well as $M_{23}$ and for the PP model we consider $M_{10}$ and $M_{13}$.

Table \ref{ParModelFit} reports the results split according to models with constant, unspecified and dynamic variance functions. In the case of constant or unspecified variance, the best performing model is again the AL model, followed by the GG model. Among distributions in the GB2 family with positive support, GG provides the best model fit according to $DIC$ with model complexity penalty while GB2 model provides the best model prediction according to $MSE$. Comparing $\bar D$s without model complexity penalty, GG and GB2 provide very similar model fit. Besides, it is clear that the PP model with only the basis function regression structure for the mean, given by a quadratic polynomial for the trend in the development year covariate, and a constant variance was not sufficient to capture all the features required. We believe that this is largely due to the fact that such a model is more suitable for heavy tailed run-off in the claims development and the Israel data clearly does not display such a feature. It is therefore expected that such a heavy tailed quantile regression model will not perform as well for this data.
When the variance is also modeled, the AL model is clearly significantly better than all the other models considered, again making $M_{23}$ with AL model optimal compared to all choices. Since, the PP model is shown to be not suitable for this data, we will consider analyses going forward with only the GB2 and AL models. \par \vspace{-5mm}
\begin{table}[H]
\caption{Parameter estimates and model fit measures for models with various distributions} \par \vspace{2mm} \hspace{10mm}
\begin{tabular} {|l|r|r|r|r|c|c|c|c|} \hline	\label{ParModelFit}
Models	 &	~~~$DIC$~~~	&	~~~$\bar{D}$~~~ &	~~~$\hat{D}$~~~	&	~$MSE$~	&	~$a$~ &	 ~$p$~ & ~$q$~ & $\sigma^2$~  \\ \hline
\multicolumn{9}{|c|}{Quantile Regression: Unspecified Variance Function}\\ \hline
$M_{2\cdot}$ Gamma &	3064.50 &	3028.93 & 2993.36 & 537.82 & 1 & 1.87 & $\infty$ & - \\
$M_{2\cdot}$ GG    &  2707.42 &	2932.97 & 3158.52 & 582.78 & 33.22 & 0.08 & $\infty$ & - \\
$M_{2\cdot}$ GB2   &	3002.82 &	2964.60 & 2926.37 & 526.65 & -7.94 & 1.78 & 0.17 & - \\ \hline
\multicolumn{9}{|c|}{Quantile Regression: Constant Variance Function}\\ \hline
$M_{10}$ PP    &	3272.14 &	1021.71 & 1230.01 & 1132.12& - & - & - & 14.15   \\
$M_{20}$ AL    &  50.94	  &120.17   &189.40   &	1015.71& - & 0.80 & - & 0.02 	 \\ \hline
\multicolumn{9}{|c|}{Quantile Regression: Non-Constant Variance Function}\\ \hline
$M_{13}$ PP    &	1502.19 &	1906.49 & 2310.98 & 923.00 & - & - & - & 9.10   \\
$M_{23}$ AL    & -20.81   & 24.91   & 70.63   &	850.10 & - & 0.75 & - & 0.17 	 \\ \hline
\end{tabular}
\end{table}

\begin{figure}[H]
\caption{\sf Standardized residual plot} \hspace{10mm}
\includegraphics[height=7cm,width=14cm]{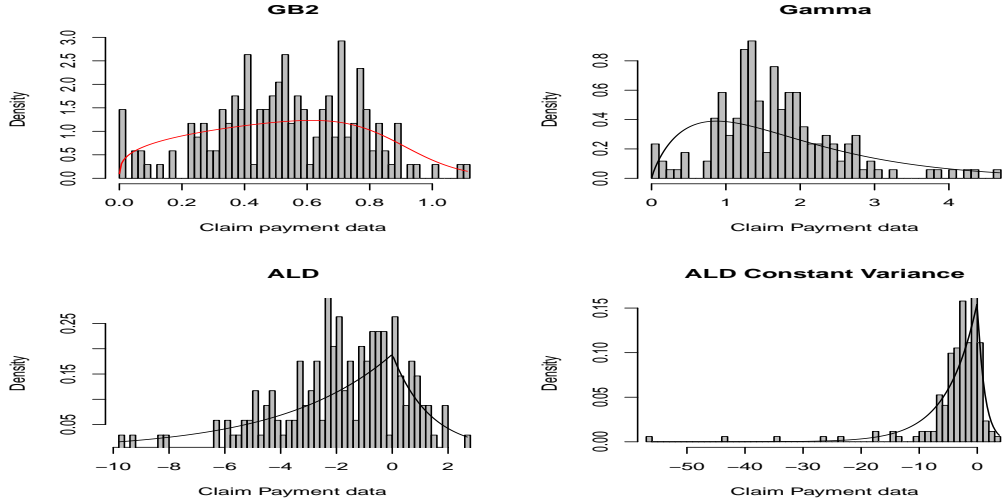} \par
\label{ResidPlot}
\end{figure}

Next, we compare the standardized residuals for the GB2, Gamma and GG models under structure in $M_{2\cdot}$ against the best fitting AL model, that is $M_{23}$ with $\hat p=0.75$. We first assess how well these models perform in sample, by looking at the following fitted model densities, versus the histograms of standardized residuals, displayed in Figure \ref{ResidPlot}. This plot shows that $M_{2\cdot}$ with GB2 distribution and $M_{23}$ with AL distribution and $\hat p=0.75$ provide good fit to the standardized residuals whereas gamma distribution provides the worst fit.


\begin{figure}[H]
\caption{\sf Percentiles of fitted losses in the upper triangle using GB2 family and AL distributions} \par \vspace{-6mm} \hspace{30mm}
\includegraphics[height=7cm,width=10cm]{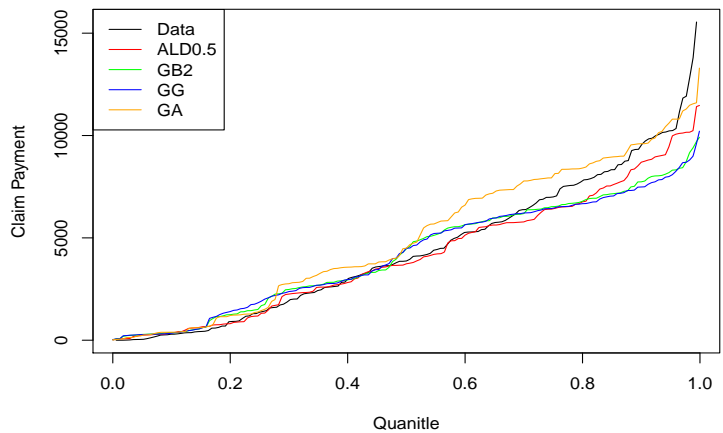} \par
\label{Percentile}
\end{figure}

Then, for out of sample analysis we display in Figure \ref{Percentile} the median predicted total claim reserve under the GB2 and AL ($p=0.5$) models. To compare these models for the out-of sample predictions we compare fitted losses of the four models by plotting $\widehat{Y}^{(\tt p)}$ against the percentile {\tt p} where $\widehat{Y}^{(\tt p)}$ refers to the {\tt p}-th percentile of all $\widehat{Y}_{ij}=\mu_{ij}$ in the upper triangle arranged in ascending order. We can see that the fitted losses using AL model are closest to the observed losses, GG and GB2 models provide very similar fitted losses and gamma model provides the poorest fit.  \par \vspace{-2mm}

\par \vspace{-7mm}
\begin{table}[H]
\caption{Selected percentiles of fitted losses in the upper triangle using GB2 and AL models} \label{percentile} \vspace{2mm} \hspace{20mm}
\begin{tabular} {|l|c|c|c|c|c|} \hline \label{PPredQuan}
Models	 &	~~~$0.30$~~~	 &	~~~$0.50$~~~	 &	~~~$0.75$~~~	 &	~~~$0.90$~~~ &	~~~$0.95$~~~	 \\\hline
Observed      &	1,985	&	3,871	&	6,990  &	9,327  &   10,200\\
$M_{2\cdot}$ Gamma 		&	 2,760	&	4,496	& 8,036	   & 9,600	  &  10,700 \\
$M_{2\cdot}$ GG 	&	2,378	&	4,498	&	6,451   &	7,486  &  8,040  \\
$M_{2\cdot}$ GB2	&	2,480 	&	4,463	&6,526	 &	7,737	& 8,247	 \\
$M_{23}$ AL ($p=0.5$) &	2,255	&	3,734	&	6,422   &	8,696  & 9,715   \\ \hline
\end{tabular} \end{table} \par \vspace{-2mm}

\begin{figure}[H]
\caption{\sf Boxplots of predicted quantile in the upper triangle using GB2 family and AL distributions} \hspace{20mm}
\includegraphics[height=8cm,width=12cm]{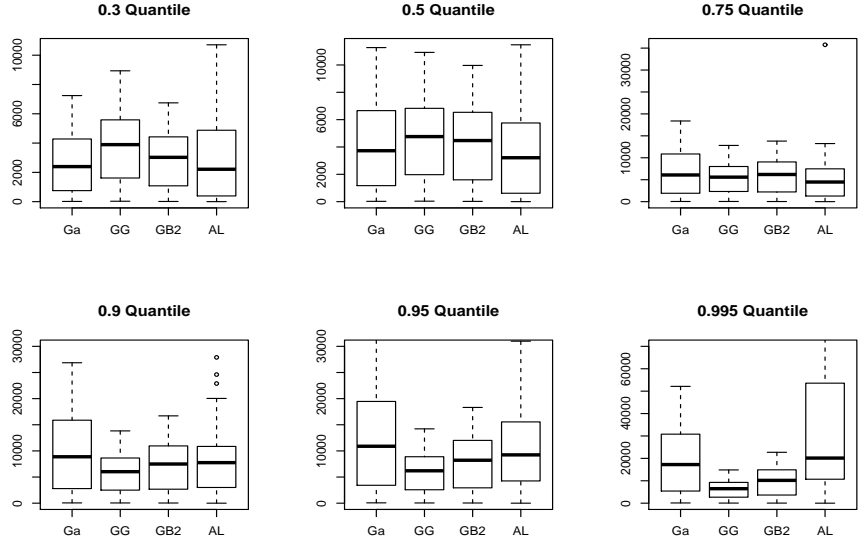} \par
\label{Boxplot}
\end{figure}

Table \ref{PPredQuan} reports the observed and fitted loss $\widehat{Y}^{(\tt p)}$ for {\tt p} = 0.3, 0.5, 0.75, 0.9 and 0.95 using the four models. As the model assessments show adequate model fits, we apply the models to predict losses at different quantile levels. Figure \ref{Boxplot} presents boxplots of quantiles $Q_Y(u| \bm x_{ij})$ for losses in each cell of the upper triangle for a given quantile level $u$ and model. Comparing across models, the boxplots for AL model have the heaviest right tails and the ranges of boxplots differ more at higher quantile level. In particular, the ranges for gamma and AL models increase much faster across quantile levels than the GG and GB2 models.

\par
\begin{figure}[H]
\caption{\sf Percentiles of predicted quantiles in the upper triangle using GB2 and AL models} \hspace{20mm}
\includegraphics[height=8cm,width=12cm]{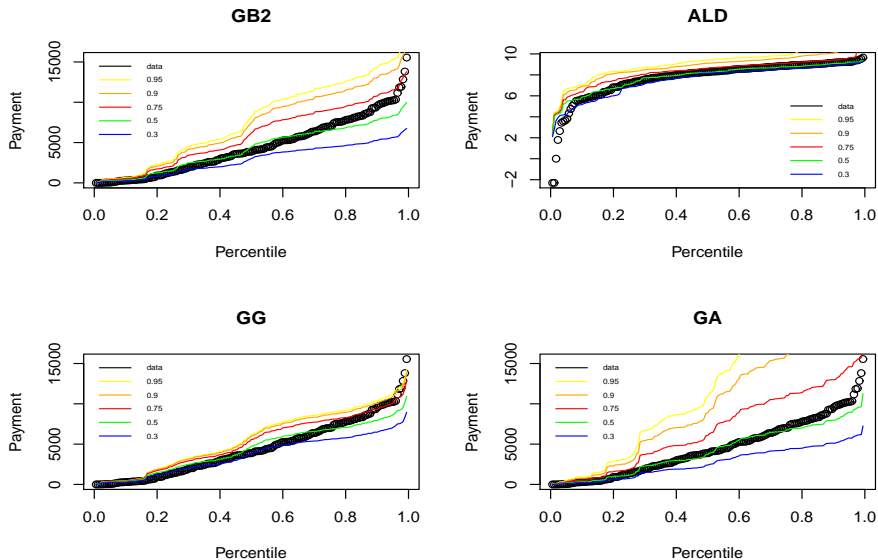} \par
\label{QPredQuan}
\end{figure}

 \vspace{-5mm} \noindent
These features can also be observed in Figure \ref{QPredQuan} which plot quantiles $Q_Y(u| \bm x_{ij})$ in each boxplot in ascending order. This is similar to Figure \ref{Percentile} but the percentile of quantiles $Q_Y(u| \bm x_{ij})^{(\tt p)}$ instead of fitted $\widehat{Y}_{ij}^{(\tt p)}$ is plotted against the percentile {\tt p}. Each line in Figure \ref{QPredQuan} corresponds to a quantile level $u=0.3, 0.5, 0.75, 0.9$ and 0.95. These so called empirical quantile lines are dense for GG model, sparse for gamma model and moderate for GB2 model indicating that GB2 distribution provides quantile estimates which can reasonably cover the observed losses across percentile {\tt p} when the quantile level $u$ gradually increases. We also remark that the empirical quantiles for AL model in the log scale are convex rather than concave and are more dense because of the log transformation.
\par \vspace{-2mm}

\begin{figure}[H]
\caption{\sf Quantile functions using GB2 family and AL distributions} \hspace{30mm}
\includegraphics[height=6cm,width=10cm]{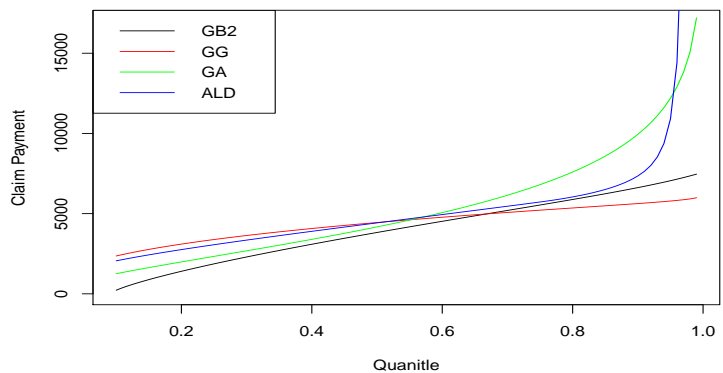} \par
\label{QuanFcnMean}
\end{figure}

Then Figure \ref{QuanFcnMean} plots the quantile functions $Q_Y(u| \bm x_o)$ across quantile levels $u \in (0,1)$ using (\ref{QuanFcnGB2}) for gamma, GG and GB2 in the GB2 family of distributions and $\exp(Q_{Y^*}(u| \bm x_o))$ in (\ref{QuanFcnPar}) where $Q_{\epsilon^*}(u)=F^{-1}_{z^*}(u)$ is given by (\ref{InvCdfAL}) for AL distribution. Note that the mean $\mu$ in $Q_Y(u| \bm x_o)$ or $\mu^*$ in $\exp(Q_{Y^*}(u| \bm x_o))$ is given by the average of $\exp(\mu_{ij}^*)$ or $\mu_{ij}^*$ over risk cells in the upper triangle.
Again AL distribution has the heaviest right tail because of the log transformation.

We further adopt these models to calculate the outstanding reserves (OR) as reported in Table \ref{PredOSQuan} using the conditional predictive posterior approach in (\ref{QuantileForTotal}) where the conditional total reserve posterior quantile function is adopted for the case of light tailed run-off in the claim process because the claim distribution was shown to be light tailed in the previous analyses.
Under the Solvency II framework, insurers will have to establish technical provisions to
cover future claims expected from policyholders. Insurers must also have available financial resources sufficient to cover
both a minimum capital requirement and a SCR. The SCR is based on a VaR measure calibrated to a 99.5 percent confidence level over a one-year time horizon. Results in Table \ref{PredOSQuan} show that the OR projection increases gradually up to 95 percentile quantile levels but increases dramatically at 99.5 percentile.

\begin{table}[H]
\caption{Outstanding reserves at different quantile levels using GB2 family and AL distributions} \par \vspace{2mm} \hspace{10mm}
\begin{tabular} {|l|c|c|c|c|c|c|} \hline \label{PredOSQuan}									
Models	 &	~~~$0.30$~~~	 &	~~~$0.50$~~~	 &	~~~$0.75$~~~	 &	~~~$0.90$~~~ &	~~~$0.95$~~~ &	 ~~~$0.995$~~~	 \\\hline
$M_{2\cdot}$ Gamma 	&	127,816	& 198,907	& 324,515	 & 474,073	& 581,302 & 920,142\\
$M_{2\cdot}$ GG 	&	203,207	& 248,409	& 291,457	 & 314,482	& 323,346 & 337,658\\
$M_{2\cdot}$ GB2	&   152,315	& 225,017	& 311,625	 & 377,154	& 413,525 &	512,731\\
$M_{23}$ AL 	&	145,031	& 176,926	& 314,454	 & 435,402	& 462,980 & 560,430\\ \hline
\end{tabular}
\end{table}

\section{RISK MARGIN: AUSTRALIAN CASE STUDY} \par \vspace{-4mm} \noindent

In general the guidance on calculation of risk margin by regulators leaves flexibility in the practical modelling approach adopted by practitioners. There are a few popular approaches considered in practice, some of which involve a degree of expert opinion. In this section we aim to consider only approaches based on statistical models and in particular percentile and quantile based methods. In this context the standard practice is to consider the reserve estimate and then try to quantify the uncertainty associated with the reserve estimator. This uncertainty is typically measured via a standard error, which is utilized to adjust the reserve. Traditionally, if a loss distribution produces an estimator for the reserve which admits a normal distribution (approximately under a central limit theorem result), then setting the risk margin to equal the sample estimator for the reserve plus 0.675 times the sample estimators standard deviation would result in risk margins calibrated to approximately the 75th percentile. Note, whilst the total loss distribution may not have finite second moment if a heavy tailed run-off is present, the variance of the sample estimator for the distribution of the reserve will always be well defined. It should be noted that this method suffers from drawbacks as there is both an influential judgment in determining the appropriate multiple, especially when the normality assumption is not present due to sample estimators distribution being skewed.

Alternatively, one may utilize the quantile regression model obtained for the total loss distribution. There are two basic ways this may be achieved, for instance one could take instead of a mean reserve, a quantile based reserve. This could be via a risk measure such as VaR which represents a tail quantile of the total loss distribution at say 99.95\%, in which case one may judge that a conservative measure of reserve is obtained from such a tail measure and so no additional risk margin is required. This is standard in banking regulations such as Basel II/III and being considered in insurance regulations.

Alternatively, one may take a central measure as the reserve such as the median of the total loss distribution and make a risk margin adjustment based on the tail quantile of the total loss distribution at say 75\% (as is considered in practice).

Thirdly, if the traditionally utilized estimate of reserve based on the mean of the loss distribution is considered, then two scenarios may arise if one uses the risk margin adjustment based on the tail quantile of the total loss distribution at say 75\%. In this case the estimated mean reserve could be below the desired risk margin quantile level of the total loss distribution, in which case it may be reasonable to make no further adjustment if the risk margin is already at a tail quantile such as 75\%. Alternatively, if the estimated mean reserve is below the desired risk margin quantile level of the total loss distribution, then the difference would be the resulting risk margin.

\begin{figure}[H]
\caption{\sf Traditional method (upper) versus proposed method (lower)}  \hspace{30mm}
\includegraphics[height=6cm,width=12cm]{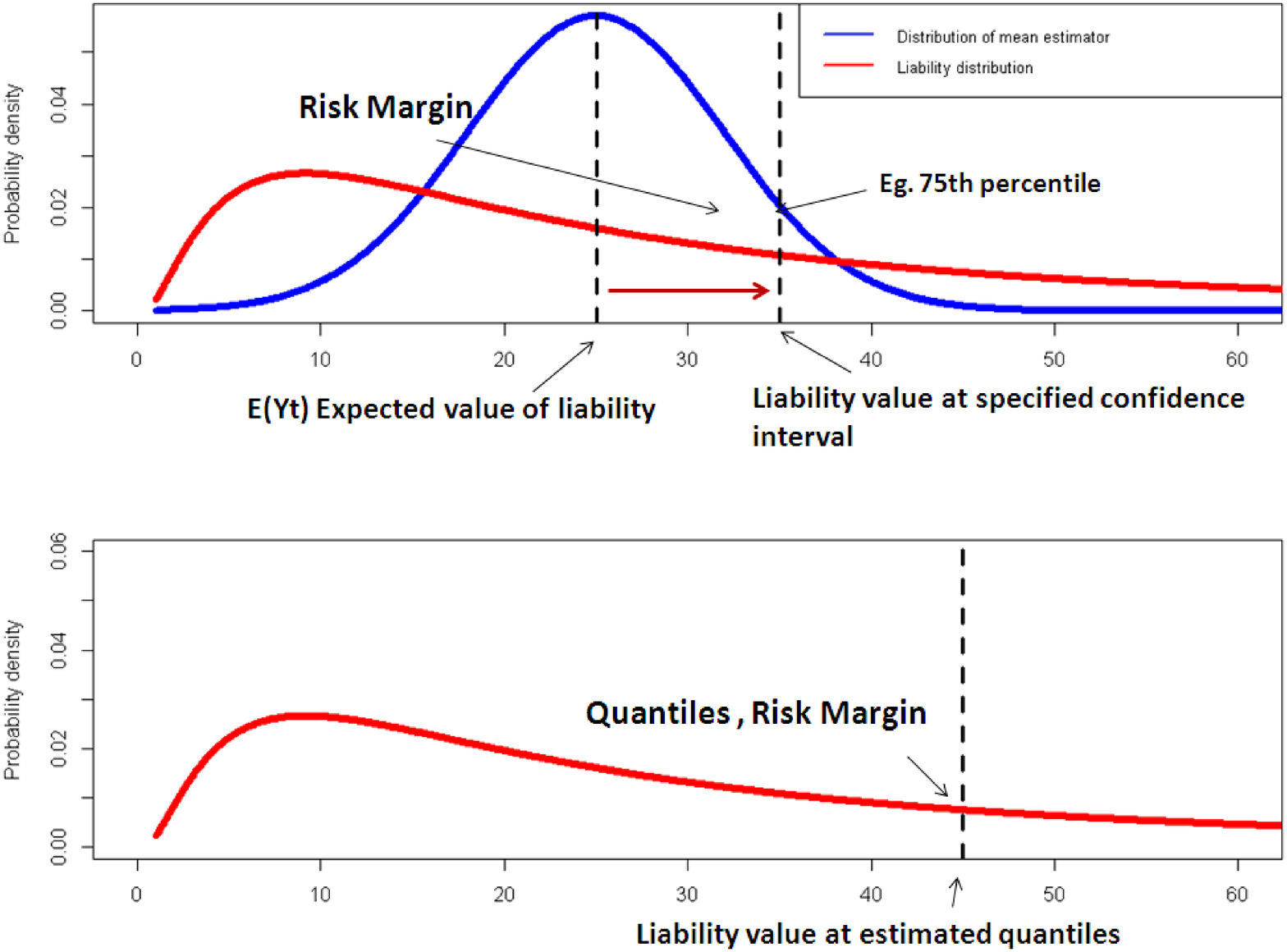}

\label{D1}
\end{figure}

In this section, we are going to extend the best model, model $M_{23}$ with AL distribution, in the previous sections to model risk margin statistically. To achieve this, we generalise the AL distribution to model the shape parameter $p$ via the following regression $p_{i}=\phi_0+\phi_i$ where $\phi_0$ is the intercept and $\phi_i$ denotes accident year effect. Accident year effect is chosen because risk capital allocation is by accident years. It is worth noting an important assumption which are stated as underlying this method: actual outstanding claim payments are assumed to be uncorrelated between accident years. Therefore, the estimated shape parameter $p$, which presents quantile in AL distribution, and also infers risk margin in the percentile method, is an applicable risk margin estimate for outstanding claims payments. The difference between our proposed method and the traditional method is also demonstrated in Figure \ref{D1}.

The data that we used to demonstrate our model is the amount of payments for all the compulsory third party (CTP) policies in Queensland (QLD) as of June 2008. CTP insurance policy covers risk that would be referred to as Auto Bodily Injury in the U.S. and Motor Bodily Injury in the U.K.. The data are in the units of millions summarized by accident and development quarters covering periods from December 2002 to June 2008. It contains 276 observations over 23 accident quarters. In order to remove the influence of inflation for reserving purposes, we utilize the average weekly earning index from the Australian Bureau of Statistics (ABS) to inflate all the values to December 2008 dollars. Hence, the data used in this analysis represents the inflated cumulative payment for QLD CTP portfolio as reported in Figure \ref{QLDCTPPayment} in appendix I.


To review features of the data, Figure \ref{QLDCTPVar} plots the observed variance across accident year on original and log scale. It shows that the variance fluctuates a lot across accident year on the original scale but displays a sharp drop on the log scale.
Figure \ref{QLDCTPSkew} shows that the skewness are mostly negative on the original and log scales. The overall skewness of the data is 0.61 and that on a log scale is -1.08. Trend of skewness reveals a sharp drop at the start and then it fluctuates across accident years for data on the original scale but increases monotony for data on the log scale.
These changes confirm the necessity of adopting dynamic variance and skewness in modelling the data.

Among choices of distributions, the AL distribution allow flexibility in modelling variance and skewness through modelling directly the scale and shape parameters $\sigma^2$ and $p$ respectively.
Furthermore, in the context of nonparametric regression using AL as a proxy distribution for model implementation, $p$ indicates the quantile level of a model which corresponds to risk margin in loss reserving. In the analysis of QLD CTP data, we adopt the ANOVA type model ($M_{23}$) for the mean and variance as it has been shown to provide the best model performance.
We further propose modelling the risk margin $p$ as a linear function of accident year.
One reason is that as accident year increases, there are more uncertainty involved in estimating the reserves; hence it is an important factor in risk margin estimation. This model is called $M_{23'}$ in the Appendix.

\begin{figure}[H]
\caption{\sf Observed variance of QLD CTP payment data by accident year} \hspace{30mm}
\includegraphics[height=5cm,width=10cm]{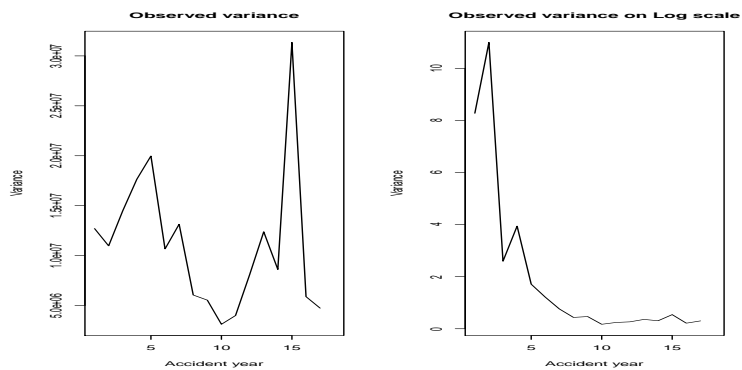} \par
\label{QLDCTPVar}
\end{figure}

\begin{figure}[H]
\caption{\sf Observed skewness of QLD CTP payment data by accident year} \hspace{30mm}
\includegraphics[height=5cm,width=10cm]{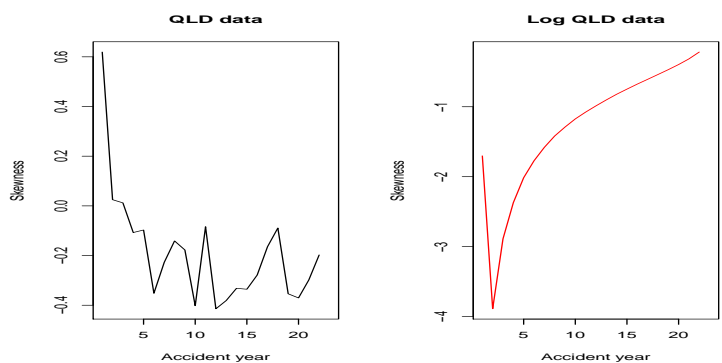} \par
\label{QLDCTPSkew}
\end{figure}




Then $M_{23'}$ with dynamic variance and skewness is compared to two models, $M_{20}$ with constant variance and skewness and $M_{23}$ with just dynamic variance in Table \ref{RiskMargin}. Although $M_{20}$ outperform $M_{23'}$ according to $DIC$, $M_{23'}$ provides the best model fit according to $\bar D$ which measures model fit alone, discounting model complexity penalty. As our aim is to provide the most accurate risk margin estimates, we adopt $M_{23'}$ in the subsequent risk margin analysis.
From a modelling perspective, it reconciles with our risk margin estimation approach.

\begin{table}[H]
\caption{Parameter estimates and model fit measures for ANOVA models using QLD CTP payment data} \par \vspace{2mm} 	
\begin{tabular} {|l|c|c|c|c|c|c|c|} \hline \label{RiskMargin}
Models	 &	~~~$DIC$~~~	&	~$\bar{D}$~ &	~$\hat{D}$~	& $E(Y)$  & $Var(Y)$ & $S(Y)$	 \\\hline
$M_{20}$ Constant variance \& skewness &	-322.55 &	-215.65 & -108.75 &  4.33 &  0.008 &  -0.28 \\
$M_{23}$ Dynamic variance &	-311.36 &	-197.71 & -84.06 & 7.67 & 0.22 & -0.57\\
$M_{23'}$ Dynamic variance \& skewness &	-255.03 &	-229.46 & -203.90 & 4.77 &  0.10 &  -0.18\\\hline
\end{tabular}
\end{table}

\begin{figure}[H]
\caption{\sf Change of $p$ across accident year using $M_{23'}$ for risk margin analysis}  \hspace{30mm}
\includegraphics[height=5cm,width=10cm]{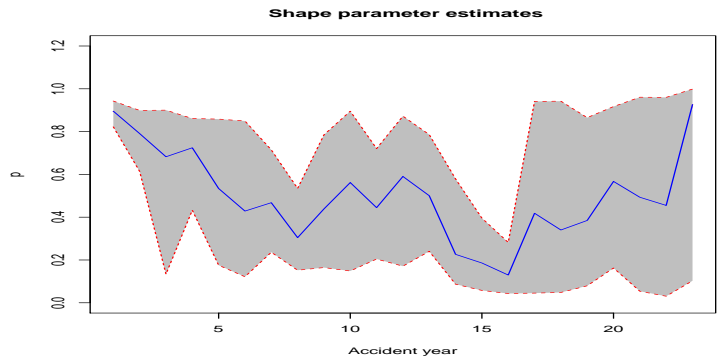} \par
\label{M20}
\end{figure}

\begin{figure}[H]
\caption{\sf Estimated variance and skewness in $M_{23'}$ for risk margin analysis}  \hspace{30mm}
\includegraphics[height=5cm,width=10cm]{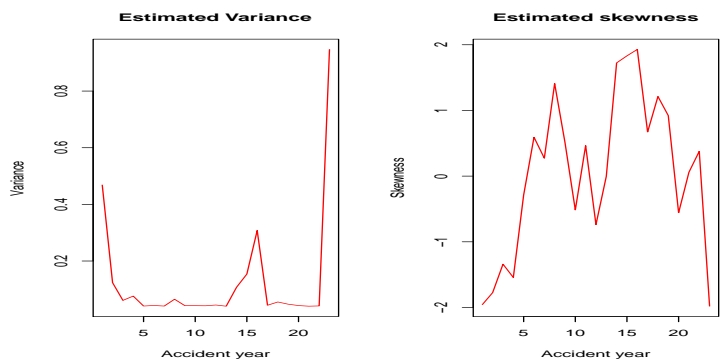} \par
\label{M200}
\end{figure}

Figure \ref{M20} demonstrates how the estimated risk margin $\hat p_{i}$ changes across accident years, superimposed with its creditable interval. Figure \ref{M200} displays the corresponding changes in estimated variance and skewness using the variance and skewness equations in (\ref{ALVar}) and (\ref{ALSkew}) respectively. The risk margin $\hat p$ starts at 0.895 at accident year 1 when the variance is quite high. Afterwards, it decreases gradually to 0.439 in accident year 8 when the variance is much smaller. From accident year 17 onwards, the risk margin increases again when the variance is large and there are more development years ahead.
In actuarial practice, the calculation of the risk margin is often not based on a sound model but various simplified methods are
used. This approach enables us to calculate a risk margin for non-life insurance run-off liabilities
in a mathematically consistent way, and provides reasonable risk margin estimates.

\normalsize\section{\normalsize CONCLUSION} \par \vspace{-4mm} \noindent
We have applied the quantile regression model to estimate loss reserve and risk margin.
Quantile regression reveals relationships between responses at the upper or lower quantiles,
which is of significant interest in estimating risk margin and VaR in insurance and finance applications. Compared to mean regression, it is more robust to heavy tailed data. We compare the performance of parametric and non-parametric quantile regression. In the parametric framework, we built five models, namely AL, PP, GB2, GG and gamma. The AL model provides the best fit. We also investigate three different regression structures, namely ANCOVA, ANOVA and Poisson-Tweedie regression. The ANOVA model performs the best in our empirical data study.

Furthermore, we adopt the best performed model, which is the AL model with ANOVA mean and variance functions, to estimate risk margin. The generalized AL model with a dynamic shape parameter $p$ provides us a mathematically consistent way of estimating risk margin. Overall, the results of our studies indicate that this new risk margins framework offers considerable potential benefits for reserving purpose.
However, the drawback is that quantile functions may cross over particularly at extreme quantiles when data are rare. Extreme quantile may not be estimated precisely.
Although there is no simple solution to this problem yet, we believe it is important to be aware of this limitation when using this framework.

\ignore{
ACF for the best model M6 ALD
\begin{center}
{\sf Figure 17: ACF M6} \hspace{10mm}
\includegraphics[height=10cm,width=16cm]{acf.eps} \par
\end{center}
}
\vspace{10mm}\par

\newpage

{\normalsize APPENDIX I}

\begin{figure}[H]
\caption{\sf Israel payment data} \hspace{1mm}
\includegraphics[height=9cm, width=17cm]{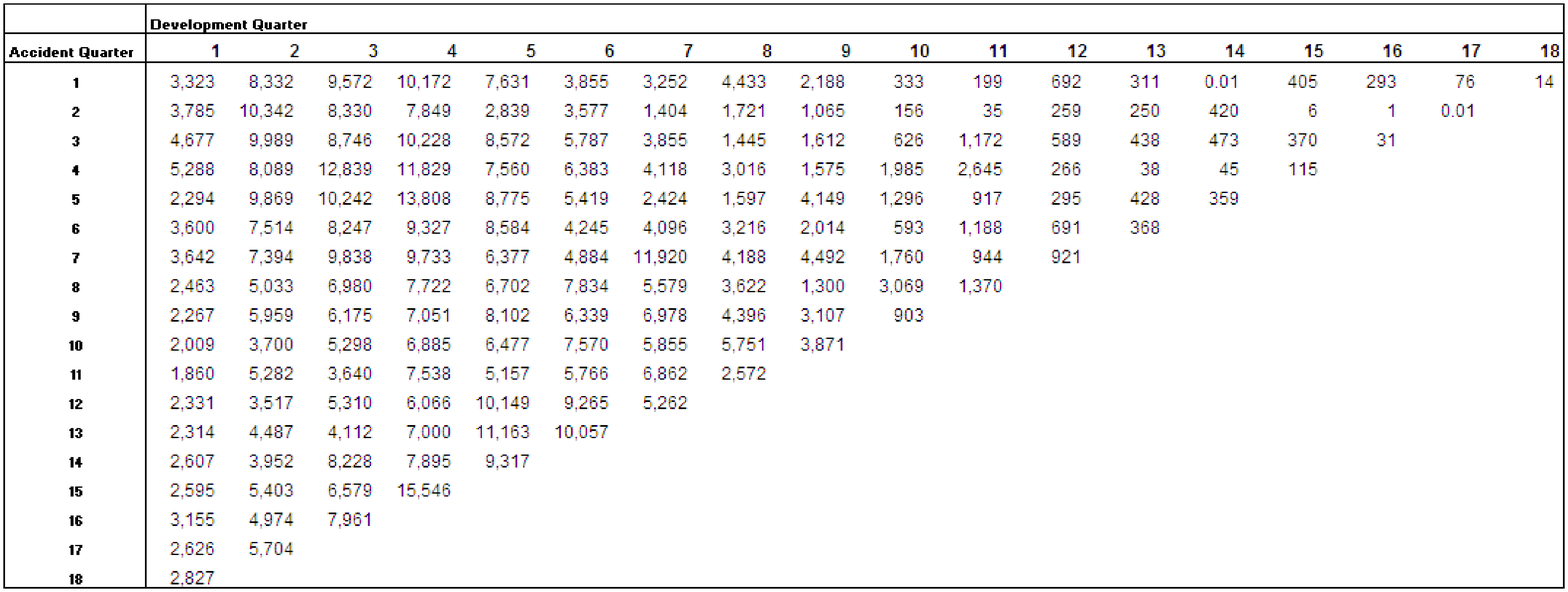}
\label{IsraelData}
\end{figure}
\par \vspace{-25mm}

\begin{figure}[H]
\caption{\sf QLD CTP payment data} \hspace{10mm}
\hspace{-10mm}\includegraphics[height=9cm,width=17cm]{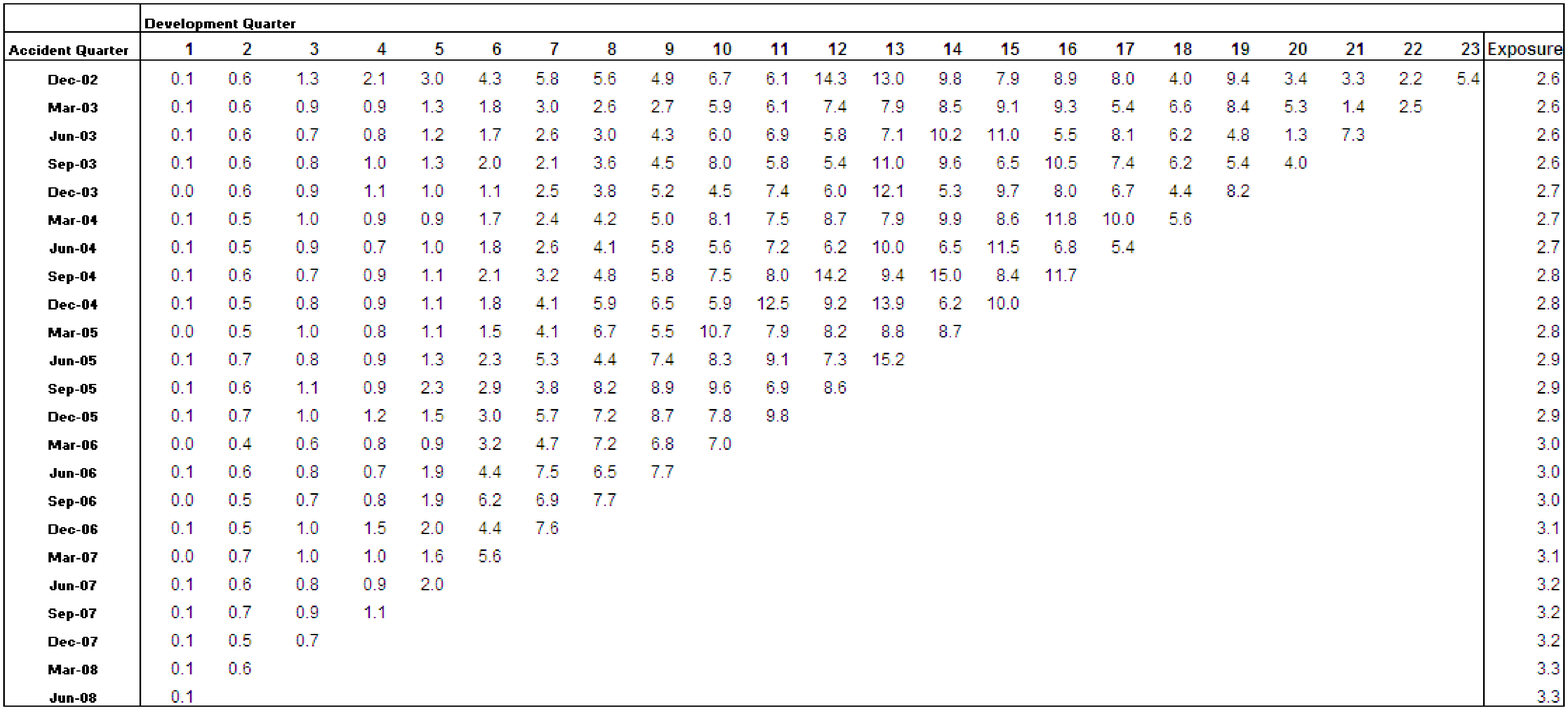}
\label{QLDCTPPayment}
\end{figure}

\begin{landscape}
{\normalsize APPENDIX II}
The following table shows the model structures considered for each regression analysis.

\begin{table}[h!]
\scalebox{0.9}{
\begin{tabular}{cccc||p{8cm}} \hline
Model Index & Model Location Structure & Model Scale Structure & Distribution Types & Model Description\\ \hline \hline
$M_{00}$ & $\mu^*_{ij} = \alpha_0 + \alpha_1 \times i + \alpha_2 \times j$ & $\sigma_{ij} = \sigma$ & AL & \textbf{Location:} Simple Additive Model (parsimonious) common trend in accident years and development years. \newline \textbf{Scale:} homoskedasticity in development years scale parameter (common across accident years).\\
$M_{10}$ & $\mu^*_{ij} = \alpha_0 + \alpha_1^S F_1(j) + \alpha_2^C F_2(j)$ & $\sigma_{ij} = \sigma$ & AL &
\textbf{Location:} Basis function regression model with trend component for development years given by Level, Slope and Curvature components (common across accident years). \newline \textbf{Scale:} homoskedasticity in development years scale parameter (common across accident years).\\
$M_{20}$ & $\mu^*_{ij} = \alpha_0 + \alpha_{1i} +{\alpha}_{2j}$ & $\sigma_{ij} = \sigma$ & AL, PP&
\textbf{Location:} Fully parameterized model with individual trend components in accident and development years. \newline \textbf{Scale:} homoskedasticity in development years scale parameter (common across accident years).\\
$M_{2\cdot}$ & $\mu^*_{ij} = \alpha_0 + \alpha_{1i} +{\alpha}_{2j}$ & Eqn 19. & GB2 &
\textbf{Location:} Fully parameterized model with individual trend components in accident and development years.
\end{tabular}
}
\caption{Model Structures in the Quantile Regressions. Note: basis function choices
$F_1(j) = \left(\frac{1 - e^{-\lambda \times j}}{\lambda \times j}\right)$,
$F_2(j) = \left(\frac{1 - e^{-\lambda \times j}}{\lambda \times j} - e^{-\lambda \times j}\right).$}
\label{TabModelStruct}
\end{table}

\begin{table}[H]
\scalebox{0.9}{
\begin{tabular}{cccc||p{8cm}} \hline
Model Index & Model Location Structure & Model Scale Structure & Distribution Types & Model Description\\ \hline \hline
$M_{21}$ & $\mu^*_{ij} = \alpha_0 + \alpha_{1i} + {\alpha}_{2j}$ & $\sigma_{ij} = \beta_0 + \beta_{1i}$ & AL &
\textbf{Location:} Fully parameterized model with individual trend components in accident and development years. \newline \textbf{Scale:} heteroskedasticity in accident years with common variance over development years scale parameter.\\
$M_{22}$ & $\mu^*_{ij} = \alpha_0 + \alpha_{1i} + {\alpha}_{2j}$ & $\sigma_{ij} = \beta_0 + {\beta}_{2j}$ & AL &
\textbf{Location:} Fully parameterized model with individual trend components in accident and development years. \newline \textbf{Scale:} heteroskedasticity in development years with common variance over accident years scale parameter.\\
$M_{23}$ & $\mu^*_{ij} = \alpha_0 + \alpha_{1i} + {\alpha}_{2j}$ & $\sigma_{ij} = \beta_0 + \beta_{1i} + {\beta}_{2j}$ & AL &
\textbf{Location:} Fully parameterized model with individual trend components in accident and development years. \newline \textbf{Scale:} heteroskedasticity in development and accident years scale parameter.\\
$M_{23'}$ & $\mu^*_{ij} = \alpha_0 + \alpha_{1i} + {\alpha}_{2j}$ & $\sigma_{ij} = \beta_0 + \beta_{1i} + {\beta}_{2j}$ \hspace{5mm} $p = \phi_0 + \phi_{1i}$ & AL &
\textbf{Location:} Fully parameterized model with individual trend components in accident and development years. \newline \textbf{Scale:} homoskedasticity in scale parameter and shape parameter $p$ (quantile level) has trend in the accident years (common across all development years).\\
$M_{30}$ & $\mu^*_{ij} = \alpha_{0,u} + \alpha_{1i,u} + {\alpha}_{2j,u}$ & $\sigma_{ij} = \sigma$ \hspace{5mm}  & AL as proxy &
\textbf{Location:} Nonparameterized model with individual trend components in accident and development years. \newline \textbf{Scale:} not defined in the model.\\
\end{tabular}
}
\caption{Model Structures in the Quantile Regressions.}
\label{TabModelStruct}
\end{table}

\end{landscape}
\end{document}